\documentclass[reprint,amsmath,amssymb,floatfix,twocolumn,superscriptaddress,aps,pre]{revtex4-2}
        \usepackage{float}
        \usepackage{graphicx}
        \usepackage{amsmath}
        \usepackage{amssymb}
        
        \usepackage{float}
        \usepackage{physics}
        \usepackage{color}
        \usepackage{xcolor}
        \usepackage{hyperref}
        \usepackage[utf8]{inputenc}

\begin{document}
       % \title{ Flocking and Clustering Behavior of Active Particles Mediated via  Velocity Alignment and Visual Perception}

       \title{From Global Flocking to Local Clustering: Interplay between Velocity Alignment and Visual Perception of Active Particles}

        \author{Mohit Gaur}
        \affiliation{Department of Physics and Astrophysics, University of Delhi, Delhi- 110089}
        \author{Arnab Saha}\email[]{sahaarn@gmail.com}
        \affiliation{Department of Physics, University of Calcutta, Kolkata- 110007}
        \affiliation{Laboratoire de Physique Théorique et Modélisation, UMR 8089, CY Cergy Paris Université, 95302 Cergy-Pontoise, France}
        \author{Subhajit Paul}\email[]{ Communicating author: spaul@physics.du.ac.in }
        \affiliation{Department of Physics and Astrophysics, University of Delhi, Delhi- 110089}

        \date{\today}
        
\begin{abstract}
Collective behavior in biological systems was first captured by the Vicsek model \cite{vicsek-1995}, in which particles align their velocities in the average direction of neighbors, leading to coherent motion and showing an order-disorder transition. However, in many complex environments, the interactions are non-reciprocal, lacking an  action-reaction symmetry. Using framework of the Vicsek model, we implement non-reciprocity by restricting interactions to neighbors located inside a finite vision cone, for a particle by limiting its set of interacting neighbors which fall within a vision-cone, providing a  minimal description for cognitive perception. Using detailed numerical simulations, we explore the  clustering and flocking behavior due to competition between noise and limited visual perception in the presence of alignment interaction. For low noise, with reduction in the vision angle the system shows transition from a global coherent motion  to locally ordered small-sized clusters. This behavior is confirmed through the steady-state distributions of velocity components and their fluctuation relative to the global mean. This is also characterized using a polar order-parameter and a two-point velocity correlation function. Interestingly, at small vision angles, particles exhibit strong short-range correlations within clusters even in the absence of any global coherence. Time-evolution of the related correlation functions illustrate the pathways towards the emergence of such structures. The time dependence of the average cluster size and the length-scale calculated from the two-point velocity correlation show scaling behavior and indicate that the emergence of density field clustering is a consequence of the velocity-field coherence. Any kind of ordering and clustering disappear in the limit of high noise and low vision-angle regime. 
\end{abstract}

        \maketitle
        
        \section{ Introduction}
Collective behavior and self-organization of living objects is quite natural across different length scales in biological systems and has been studied extensively within ``active'' matter framework \cite{vicsek-1995,ramaswamy_mechstat_2010,collectivemotion,marchetti_hydro_active13,bechinger2016active,chate2020dry,birdflocksasCM,Czir_k_1997_collective,Beekman2001_collective_motion_ants,Czirok1996_collective_motion_bacteria,BECCO2006487_collective_motion_fish,Ballerini_2008_birds_collectivemotion,chate2008collective,Gliding_Bacteria,Kohler2011_ACTIN,solon2015phase,paul2021clusters,bera_solvpre_2022,shee_rmp_2025,paul2024dynamical,chate2024dynamic,patel2025crossover}. ``Active'' objects can maintain coherent motion using its internal energy depot where energy is assimilated from the environment \cite{ramaswamy_mechstat_2010,marchetti_hydro_active13,collectivemotion,birdflocksasCM}. Flocks of birds \cite{birdflocksasCM,birds_exp_collectivemotion_1,Ballerini_2008_birds_collectivemotion}, clusters of bacteria \cite{Gliding_Bacteria,Czirok1996_collective_motion_bacteria}, school of fish \cite{BECCO2006487_collective_motion_fish}, swarm of insects \cite{Beekman2001_collective_motion_ants}, cytoskeletal filaments driven by molecular motors\cite{Schaller2010},  etc. are a few well-studied examples. Also non-living systems, e.g., nematic fluids \cite{swarming_granular_nematics}, vibrobots,  \cite{coll_behaviour_inanimate_boat}, artificial synthetic colloids, vibrating granular beads, etc., exhibit similar collective behavior. Despite having different mechanism of microscopic motion, the ordered collective dynamics is quite generic in many systems \cite{collectivemotion}. %Thus, modeling flocking behavior via a universal picture appears quite challenging. 

In this regard, the first minimal model was by Vicsek et al.\cite{vicsek-1995} with a simple dynamical rule of velocity alignment among the neighbors. Depending upon the number density of the particles and the noise strength, this model exhibits an order-disorder transition accompanied by formation of traveling bands \cite{chate2008collective,vicsek-1995} and coexisting micro phase separation \cite{solon2015phase}, even in two dimensions.  To imitate flocking in various realistic situations, a few modified versions have also been studied. For example, by introducing a vector noise \cite{chate2008collective,Cambui_modern_physics_letters_2020}, particles executing Brownian motion with velocity alignment \cite{Grossmann_2012}, alignment interaction with binary species \cite{chatterjee_binaryVM_2023} as well as at low noise and low velocity of particles \cite{rubio_lowdens_2019}, turning away of particles instead of aligning \cite{flockingturningaway}, probing flocking in heterogeneous environments \cite{paul2021clusters,dutta_heteroVM_2025}, alignment interaction with or without cohesion \cite{gregoire2004onset,paul2021clusters}, flocking of active particles by predicting their position and neighbors after each reorientation \cite{barreto_2025_pre_predictive}, etc.
        %\textcolor{red}{ The fascinating thing about flocks of starlings, for instance, is not simply the order, but their collective ability to avoid predators, as it requires the birds to respond as soon as the flock changes direction.  }

Depending upon the total (system and environment) momentum conservation, active systems are usually broadly categorized as: ``Dry'', in which the total momentum is not conserved \cite{chate2020dry} and ``wet'' with momentum conservation \cite{marchetti_hydro_active13,bechinger2016active}.  A prominent example of a ``dry'' active material is active colloids \cite{zottl_abpcolloids_2023}. However, in some situations, like in dense complex media, there can be local momentum transfer between the colloid and its environment. Theoretically it can be modeled by active Brownian particles (ABP), where the particles move with constant speed along their direction of self-propulsion which changes diffusively  \cite{romanczuk2012active}. With or without any alignment interaction and in the presence of repulsive steric interaction, a system of ABPs exhibit clustering, known as motility induced phase separation (MIPS) when the activity and number density is sufficiently large \cite{MARCHETTI201634,fily2012athermal,redner2013structure,wysocki2014cooperative,stenhammar2014phase,wysocki2016propagating}. Being inherently out-of-equilibrium due to driven by stochastic noise,  momentum conservation is not applicable for active systems in general \cite{lowen2020inertial,scholz2018inertial}. Moreover, a  few recent works have explored that alignment rules are not always necessary to mimic flocking or orientational ordering \cite{shee_rmp_2025,flockingturningaway}.  %Even a torque generated by a particle while turning away from its neighbors can also lead to orientational ordering \cite{flockingturningaway}.

 %\cite{baconnier2024selfaligningpolaractivematter,PhysRevE.74.061908}       

In literature, most of the many-body models consider spherical particles with reciprocal interactions to understand spatial organizations or clustering \cite{vicsek-1995,fily2012athermal,romanczuk2012active,shaebani2020computational,shee_rmp_2025}. However, out-of-equilibrium dynamics in real systems mediated by various complex interaction forces  do not always follow the action-reaction symmetry and thus  not  derivable from a conservative potential. These are known as ``non-reciprocal'' (NR) interactions in literature \cite{fruchart2021non,saha2020scalar,loos2020irreversibility,paul2026interplay,de2022collective}. For a particle with shape asymmetry or an active particle having self-propulsion NR interactions is expected. However, in passive systems also, effective interactions can arise as NR. For example, in a system of passive  particles of different mobilities \cite{kolb_activebinarysm_2020}, hydrodynamic flows or chemical fields generated by Janus or catalytic colloids with different phoretic responses \cite{canalejo_prlchemically_2019,babak_phoreticprl_2020}, etc. can induce asymmetric inter-particle forces. NR interactions are inherent in many complex systems, e.g.,  in networks of neurons \cite{derrida_eplneural_1987}, directed interfacial growth, quorum sensing of particles \cite{speck_prequorum_2020}, epidemic spreading \cite{paoluzzi_epidemicsm_2020}, systems with mixture of  passive and active particles \cite{stenhammer_mixtureprl_2015}, Brownian particles with shape asymmetry \cite{liao2020dynamical,jayaram2020scalar,broker2024collective}, etc.   Even though models related to active polymers or filament-like objects are used to mimic elongated biological objects and their dynamics for directed motion \cite{paulsm_2022vic,paul2022activity}, however, are not fundamental to mimic asymmetry in forces exerted on it \cite{jayaram2020scalar,theers2018clustering}. As the presence of non-reciprocity in interaction fundamentally modify the single-particle (in a collection) as well as collective behavior, leading to occurrence of novel phenomena compared to equilibrium scenario \cite{fruchart2021non}.  A few recent works have explored different aspects using microscopic simulations with explicitly including non-reciprocity or using a field-theoretic approach considering terms with broken time-reversal and parity  \cite{saha2020scalar,fruchart2021non}. 
%They are responsible for the emergence of Time-Dependent Phases (TDPs), such as traveling density waves (an active self-propelled smectic phase) \cite{saha2020scalar} and chiral motion\cite{fruchart2021non}, where spontaneously broken symmetries are dynamically restored.

For any biological entity, vision is one crucial mechanism in determining the neighbors and correspondingly the dynamics. Such biological ``active'' agents, during their motion,  also follow the positions of their neighbors beside their velocities. To include non-reciprocity via these effects, interacting neighbors are considered within a cone governed by vision angle.  This approach has been explored recently to study collective behavior in a few   lattice and off-lattice models \cite{francois_visual_2019,negi2022_abp,navas2024impact,dinelli2023non,durve2018active}.  Whereas a delay in interaction without any attractive force can lead to a very condense ``drop'' with some orientational order within it, ABPs having persistent motion can lead to MIPS, worm-like aggregates or a phase of their co-existence, depending upon the vision angle \cite{negi2022_abp}. For lattice XY model with NR interaction, long-range order can emerge at low temperatures as a result of competitive effect between VC and lattice geometry, breaking the lattice symmetry \cite{Bandini_2025}. A few other works have considered NR coupling between fields of two species and showed various patterns depending upon  their composition \cite{kreienkamp2024nonreciprocal}. Whereas many of the studies focus on the wave propagation in NR media, or patterns emerging from order-parameter field, for microscopic models it is less explored.

In this work, we consider particles with Vicsek-like velocity alignment interaction, but the neighbors are identified with a vision-based sensing and within a cut-off distance. We aim to explore the competitive effect of alignment and visual perception and to disentangle their role in the self-organized pattern. For a fixed overall density, we vary the vision angle as well as the strength of the  noise. At low-noise and full vision angle the system forms a single stable cluster with coherent motion of the particles, whereas  at lower vision-angle and higher noise the clusters form and break  intermittently. The flocking is quantified by a polar order parameter and with equal time two-point velocity correlation function (VCF). For low vision angle, lower noise leads to small sized clusters with local ordering, whereas velocities of the neighboring particles remain uncorrelated for higher noise strength even with an explicit velocity alignment interaction. Emergence of global order is possible only for a full vision angle. In addition to the velocity flocking behavior, the density field clustering  quantified via the time evolution of the number of clusters or the average radius of gyration, show significant dependence on the VC of interaction. More generally, our detailed analysis provides understanding on the emergence of local or global polar order in connection with the density field clustering and show how information propagation quantified by correlations gets affected in the context of NR interaction provided by VC approach.

The rest of this paper is organized as follows. Section II describes the details of our model implementation and the simulation method. In Sec.~III, we present results from our simulations. Finally, in Sec.~IV, we provide a detailed summary along with a future perspective.

        \section{ Model and Methods}
        \subsection{Details of Model and Simulation}
\begin{figure*}
            \centering
          \includegraphics[height=6.2cm, width=17cm]{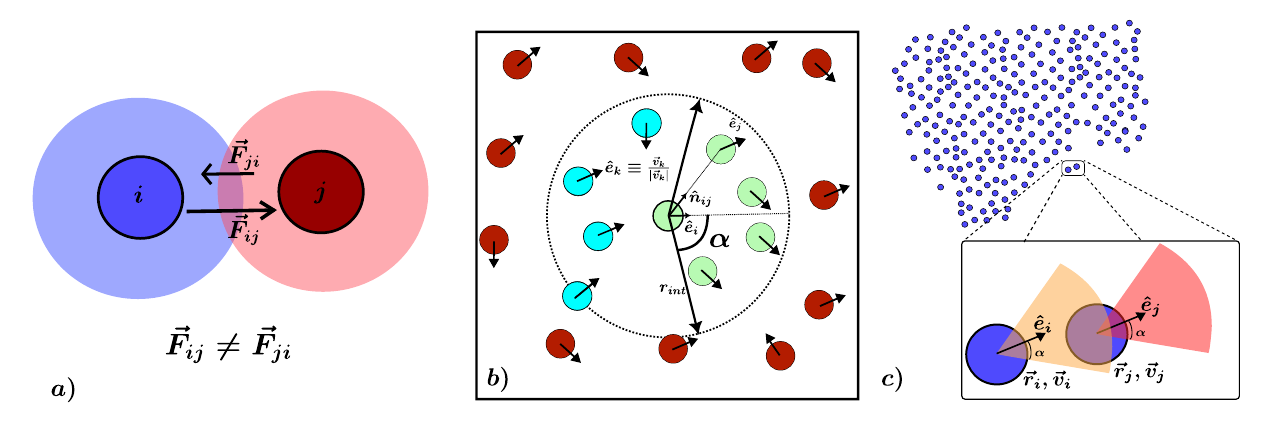}
          \caption{(a) Schematic representation showing non-reciprocal interaction between any two species, say, $i$ and $j$. Inter-particle forces $\vec{F}_{ij} \ne \vec{F}_{ij}$ accounts for such non-reciprocity. (b) Implementation of non-reciprocity via the vision cone with half-angle $\alpha$  within which $i$-th particle (position $\vec{r}_i$ and velocity $\vec{v}_{i}$) interacts with the neighbors within cut-off distance $r_{int}$ (not to scale). For clarity, interacting particles including the $i$-th one is shown by green color. Other particles (marked with cyan) within $r_{\text{int}}$ which are not within the vision cone and do not contribute to the interaction. (c) Enlarged version to show that the regions of interaction which falls within the vision cone become different for particles $i$ and $j$. Due to limitation in visual perception the interacting neighbors for a particle are always in its heading direction and thus creates anisotropy in  interaction.}\label{fig1_schematic}
        \end{figure*}
We consider $N$ particles in a box of dimension $L\times L$ with  their positions $\vec{r}_i(t)$ ($i=1, \cdots N$) randomly distributed,  and velocities $\vec{v}_i(t)=v_0 (\cos(\theta_i(t)),\sin(\theta_i(t)))$ with fixed amplitude $v_0$ and random orientations $\theta_i \in [-\pi,\pi]$. At time $t+\Delta t$, the orientation of each particle $\theta_i(t)$ is updated via \cite{vicsek-1995}
\begin{equation}
             \theta_i(t+\Delta t)= \langle \theta_i (t)\rangle_{\mathcal{N}_i(t)} +\zeta_i(t)
        \end{equation}
        where the average  $\langle \cdots \rangle$ is taken over the number of particles $n^{vc}_i(t)$ at time $t$ which are considered within a neighborhood $\mathcal{N}_i(t)$ of $i$-th particle, defined by a circular region of cut-off radius $r_{\text{int}}$ and follow the constraint of vision restriction within the angle $\alpha$ as \cite{negi2022_abp,barberis_2016} 
        \begin{equation}\label{eqn:vision_condition}
            \hat{e}_i \cdot \hat{n}_{ij} \geq \cos \alpha.
        \end{equation}
Here $\hat{e}_i\equiv(\cos \theta_i,\sin \theta_i)$ denotes the direction of velocity $\vec{v}_i$ (orientation) and  $\hat{n}_{ij}= (\Vec{r}_{i}-\Vec{r}_{j})/|\Vec{r}_{i}-\Vec{r}_{j}|$ is the unit vector in the direction joining particles $i$ and $j$, and $\alpha$ measures the half-angle of the vision cone, as seen from the schematic in Fig.~\ref{fig1_schematic}(b). A general depiction of non-reciprocity with unequal pairwise interacting forces with $\vec{F}_{ij} \ne \vec{F}_{ji}$ is shown in Fig.~\ref{fig1_schematic}(a). An enlarged version of (b) is shown in (c) where it shows that the regions of interactions for particles $i$ and $j$ (marked by their corresponding positions and velocities) are always along their heading direction and does not overlap for lower values of $\alpha$. $\zeta_i$ provides an additional fluctuation to the direction of the particle $i$. $\zeta_i$ is considered as a uniform random noise in the range $[-\eta/2,\eta/2 ]$ and delta-correlated over space and time as $\langle \zeta_i(t)\zeta_j(t')\rangle=\delta_{ij}\delta(t-t')$. Following update of $\theta_i$, the velocity of the $i$-th particle is updated as $\vec{v}_i(t+\Delta t)=v_0 (\cos \theta_i(t+\Delta t),\sin \theta_i(t+\Delta t))$. The corresponding positions $\vec{r}_i$ at $t+\Delta t$ are updated using the corresponding velocities as \cite{frenkel2023understanding}
        \begin{equation}
        \label{eqn:positionupdate}
                \Vec{r}_{i}({t+\Delta t})=\Vec{r}_{i}({\,t}) + \Vec{v}_{i}({t+\Delta t})\Delta t\,.
        \end{equation}
        In this model, particles move off-lattice and have no inter-particle interaction, i.e., particles without any finite size, can overlap, and thus the local density can be arbitrarily large. 

For all our simulations, as mentioned, we consider particles within a square box of length $L$ with periodic boundary conditions (PBC) applied in both directions. Unless otherwise mentioned, the number density    $\rho (\equiv N/L^2)=2.5$ with $L=32$ and interaction radius for the neighborhood calculation is chosen as  $r_{\text{int}}=1.0$. The integration time step $\Delta t$, and the velocity magnitude $v_0$ are chosen as $\Delta t=1.0$ and $v_0=0.01$. A few different strengths of  noise $\eta$ are considered in the range $[0.01:4.0]$. The angle of vision $\alpha$ is considered for a few discrete values between $[0,\pi]$. The choice of $\alpha\ne \pi$ gives rise to the possibility  of the emergence of non-reciprocity in our model. All of our presented data are averaged over at least $100$ independent runs.  %In addition to this, the steady-state results are also averaged over $50$ uncorrelated configurations for each run. 

%For all our simulations, as mentioned, we consider $N$  particles randomly oriented in a square box of length $L$ with periodic boundary conditions (PBC) in both the directions. Unless otherwise mentioned, the number density    $\rho(\equiv N/L^2)=2.5$ with $L=32$ and cut-off radius for neighborhood calculation as $r_{int}=1.0$. Time step of integration $\Delta t$, and velocity magnitude $v_0$ are chosen as $\Delta t=1.0$ and $v_0=0.01$. The vision-angle $\alpha$ is considered for a few discrete values between $[0,\pi]$. Choice of $\alpha\ne \pi$ is a possible origin of emergence of non-reciprocity in this case. All our presented data are averaged over at least $50$ independent trials. In addition to this, various steady-state results are also averaged over a few uncorrelated configurations for each run. 
        
        %The initial configurations for each trial are considered by randomly distributing particles on a grid, and random directions are assigned for the theta such that $\langle\theta\rangle=0$
        %since theta is zero 

        \subsection{Quantities of Interest}
        Below we describe a few quantities which are the main focus to look at in this paper.
         
        \subsubsection{Velocity and connected correlation function}
        As the dynamics is related to the velocity ordering of the particles, spatial velocity correlation $C_v(r)$ is an appropriate measure to quantify the emergence of correlation. $C_v(r,t)$ between two particles $i$ and $j$ having their velocities $\vec{v}_i$  and $\vec{v}_j$, respectively, at time $t$, is defined as 
        
        %\begin{equation} \label{vcf_defn}
        %C_{v}(r,t)= \frac{\sum_{i,j} \Vec{v}_i \cdot \Vec{v}_j \ \delta(r -r_{ij})}{\sum_{i,j} \delta(r -r_{ij})} \, -  \langle \vec{v}_i \rangle \langle \vec{v}_j \rangle
        %\Big|\sum_{i,j}^{N} \Vec{v_i} \Big|^2 
        % C_{v}(r,t)= \langle\Vec{v_{i}}. \Vec{v_{j}} \ \delta(r -r_{ij})\rangle \, - \Big|\langle  \Vec{v} \rangle\Big|^2
        %\end{equation}

        \begin{equation}\label{vcf_defn}
            C_v(r,t)= \langle \vec{v}_i(0,t) \cdot \vec{v}_j(r,t)\rangle - \langle \vec{v}_i(0,t) \rangle \langle \vec{v}_j(r,t) \rangle\,,  
        \end{equation}
        where $\vec{r}=\vec{r}_i-\vec{r}_j$ is the separation between the particles and $\langle \cdots \rangle$ defines the average over all the particles. In studies related to flocking the deviations of the velocities with respect to its global mean is also a quantity of interest. In this regard, the connected correlation function (CCF) is defined as \cite{CAVAGNAcorrelation}
        
        %\begin{equation}\label{ccf_defn}
        %C_{\delta v}(r,t)= \frac{\sum_{i,j} \delta \Vec{ v}_{i} \cdot  \delta \Vec{v}_{j} \ \delta(r -r_{ij})}{\sum_{i,j} \delta(r -r_{ij})} \, - \langle \delta \vec{v}_i \rangle \langle \delta \vec{v}_j \rangle
        % C_{\deltav}(r,t)= \langle\Vec{\delta v_{i}}. \Vec{\delta v_{j}} \ \delta(r -r_{ij})\rangle \, - \Big|\langle  \Vec{\deltav} \rangle\Big|^2
        %\end{equation}
        
        \begin{equation}\label{ccf_defn}
            C_{\delta v}(r,t)= \langle \delta \vec{v}_i(0,t) \cdot \delta \vec{v}_j(r,t)\rangle - \langle \delta \vec{v}_i(0,t) \rangle \langle \delta \vec{v}_j(r,t) \rangle\,,  
        \end{equation}
        where  $\delta \vec{v}_i = \vec{v}_i - \vec{v}_{\text{avg}}$ defines the fluctuation of the velocity of the $i$-th particle compared to the global average velocity $\vec{v}_{\text{avg}} =(\sum_{i=1}^N \vec{v}_i)/N$ calculated with all the particles in the system. CCF is particularly important as it eliminates the mean effect of the system. For example, in case of a global order, whereas the non-connected correlation or VCF should show a long-range order, CCF can provide an idea regarding to what extent the fluctuations are correlated.
        
        %$r_{ij}=|\Vec{r_{i}}-\Vec{r_{j}}|$ is the separation distance between any two given agents $\Vec{v_i}$ is the velocity of the $i^{th}$ particle, $\delta \Vec{v_i}= \Vec{v_i} -\Vec{v_{avg}}$ is the deviation of the velocity from the average. Equation \ref{eqn:vcf} takes the sum of the inner product between the velocities, divides it by the number of agents between $\Vec{r}$ and $\Vec{r} + d\Vec{r}$ and then subtracts the average of $v^{2}$. In \ref{eqn:ccf} the connected correlation function is calculated by taking the sum of  inner product of the fluctuation or the deviation of the velocity of each agent from the mean in between $\Vec{r}$ and $\Vec{r} + d\Vec{r}$.Since the fluctuations are random the second term in \ref{eqn:ccf} should be zero since the average of the random fluctuations is zero.
        
        Both velocity and the connected correlation functions depict how the different flocks in the system are related to each other or the length-scale over which the particles are aligned to each other in the system. To understand this we consider two clusters (say, $C_1$ and $C_2$) moving in different directions. However, if a few particles in $C_1$ have the same direction as the particles in $C_2$, then their velocities are correlated.  However, the velocity fluctuations which is obtained after subtracting the average value from the velocities, can be  uncorrelated or correlated over a shorter distance. Thus, the fluctuations in velocity field and related CCF can  provide a better estimate for distinguishing between a global order to a local coherent motion.
        \subsubsection{Mass of the clusters and its distribution}
        To characterize the size of the clusters, we have calculated the cluster mass distribution function \cite{peruani2006_pre,negi2022_abp}
        \begin{equation}
            %P(m_c)=\frac{1}{{\mathbb{N}}} ~m_c~p(m_c) 
                P(m_c)=\frac{1}{{{N}}} ~m_c~p(m_c)
        \end{equation}
        where $p(m_c)$ counts the number of clusters with mass $m_c$ where, $m_c$ counts the number of particles in it. The distribution function is normalized by $N$, such that $\sum_{m_c=1}^{N} P(m_c)=1$. The clusters are identified using an algorithm known as Breadth First Search, which is usually used to find the shortest path between nodes in an unweighted graph network \cite{ismail1989multidimensional}. To define a cluster, we also set a criteria for cutoff distance $r_{\text{cut}}=0.5$ and a critical size $n_{\text{crit}}=25$. Thus two particles belong to a cluster if distance between them is lower than $r_{\text{cut}}$ and the cluster is well defined if the group size is higher than $n_{\text{crit}}$. This is discussed in more detail in Appendix~\ref{Appendix_clust_id}.
        
        \begin{figure*}
        \includegraphics[width=17.5cm, height=10cm]{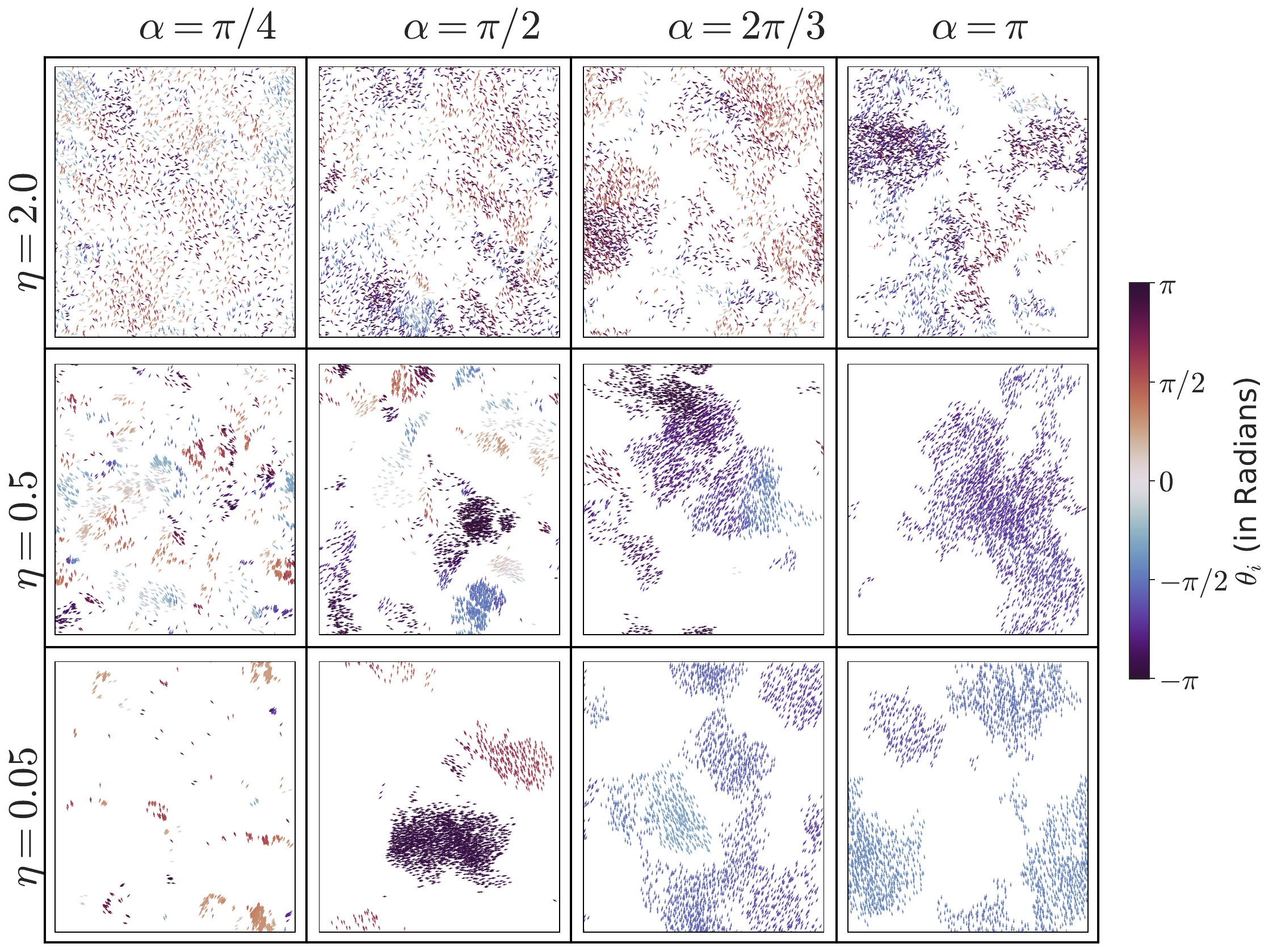}
          \caption{%Typical snapshots of the velocity field of the particles at the steady-state for different $(\alpha=\pi/4,\pi/2,2\pi/3,\pi)$ and $(\eta=0.05,0.5,1)$
          Typical steady-state snapshots of the velocity field of the particles for different values of the vision-angle $\alpha$ and noise strength $\eta$, as mentioned. Different colors (shown in the bar) identify the regions of different orientations of the particles. For $\alpha=\pi$, i.e., with full vision angle,  snapshots for different values of $\eta$ look similar to the ones for original Vicsek model.} 
          \label{snapshot_ss}
        \end{figure*} 
        \subsubsection{Radius of gyration of clusters}
        The typical size or span of a cluster can be estimated by calculating its radius of gyration ($R_g$) defined as \cite{rubinstein2003polymer}
        \begin{equation}
            R_g= \Big[\frac{1}{m_c} \sum^{m_c}_{i=1} (\vec{r}_i-\vec{r}_{cm})^2\Big]^{1/2}\,
            \label{radius_of_gyration}
        \end{equation}
        where $m_c$ is the number of particles or mass of the cluster as defined earlier, and $\vec{r}_{cm}=\big(\sum_{i=1}^{m_c} \vec{r}_i\big)/m_c$ is the center of mass of the cluster. The average length-scale of the clusters are calculated by considering $R_g$ of all the clusters present in the system. However, $R_g$ has an important limitation in accurately predicting the size of the clusters, as the particles here do not have a finite size and hence can overlap in high enough density at low enough noise. We will return to discussing this point where the results regarding the calculation of $R_g$ will be discussed. Furthermore, in the presence of PBC in the system, some part of a cluster can be present on the other side \cite{bai2008calculating} which can affect the calculation of $R_g$.  The calculation of $R_g$ for a cluster in  such a scenario  is depicted in  Appendix~\ref{Appendix_cm}.
        %\subsubsection{Adjacency matrix \textcolor{red}{needs rewriting, will do later}}
        %The  of the non-reciprocity can be done using the adjacency matrix $A$. Elements of this matrix $a_{i,j}$ can be written in a binary manner ($1$ or $0$). $a_{i,j}$ is taken as $1$ only for the neighboring particles which are both within the cut-off distance $r_{int}$ and vision cone $\alpha$.
        
        \section{ Results}
        We present our results in two sub-sections. In the first one we discuss the results for the steady state of the system with different parameters. Results related to the non-equilibrium dynamics are discussed in the second sub-section.

        \subsection{Steady State Results}
        \begin{figure}[htb]
        \hskip -0.1in
          \includegraphics[width=8.8cm, height=7cm]{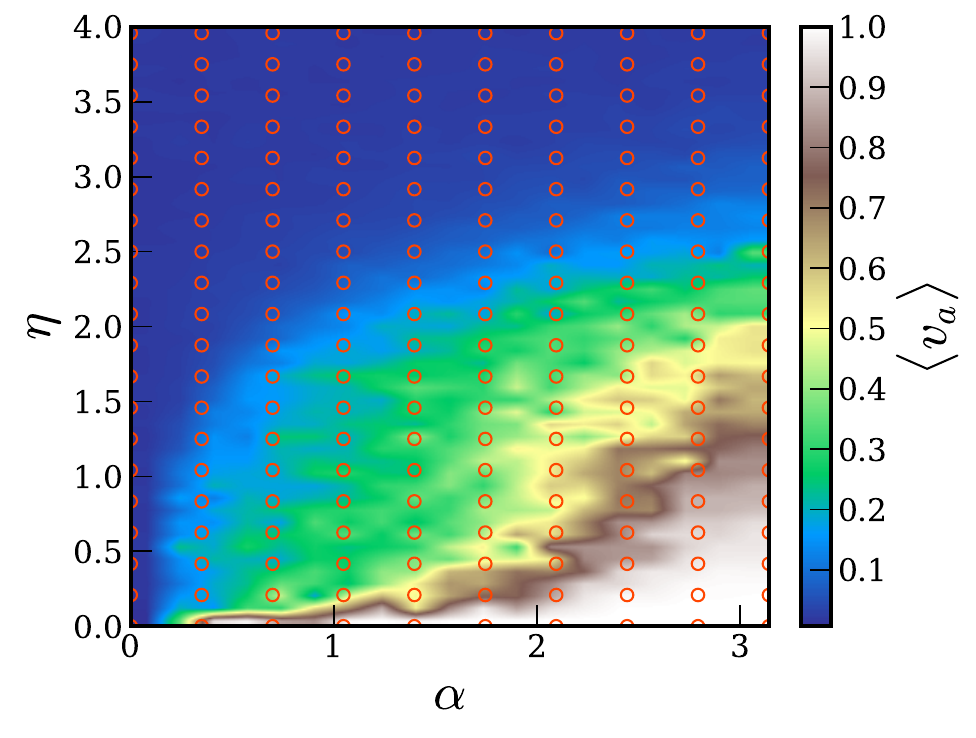}
          \caption{Heatmap of $\langle v_{a} \rangle$  in $\alpha - \eta$ parameter space. The color bar represents the magnitude of $\langle v_a\rangle$ which varies between $0$ to $1$. The points for simulations are marked by circles.}
          \label{va_heatmap}
        \end{figure} 
        We begin by showing the typical steady-state snapshots of the system for different combinations of vision angle $\alpha$ and noise $\eta$ in Fig.~\ref{snapshot_ss}, at fixed $\rho=2.5$. For $\alpha=\pi$, i.e., the standard Vicsek model {\cite{vicsek-1995}}, depending on different choices of $\eta$, one observes the particles can organize themselves to large clusters and coherent motion. However, with decreasing $\alpha$ the formation of such large clusters as well as global coherent motion becomes less probable. For our lowest choice $\alpha=\pi/4$,  with lower noise $\eta=0.05$, small dense and locally coherent clusters are possible, however for $\eta=2.0$ the particles remain fully incoherent and do not form any cluster. The absence of any typical particle size makes these clusters dense for smaller values of noise (as observed for $\alpha=\pi/4$ with $\eta=0.05$ and $0.5$). The choice of these parameters gives a qualitative idea about the interplay between noise $\eta$ and the vision angle $\alpha$ at a fixed particle density. Thus, global ordering seems possible only with $\alpha=\pi$. 
        %Small $\alpha$ and small $\eta$ promote the formation of dense clusters and local ordering, which, in turn, renders velocities uncorrelated beyond these clusters. 
        \begin{figure}[t!]
        \hskip -0.25in
            \includegraphics[width=8cm, height=10cm]{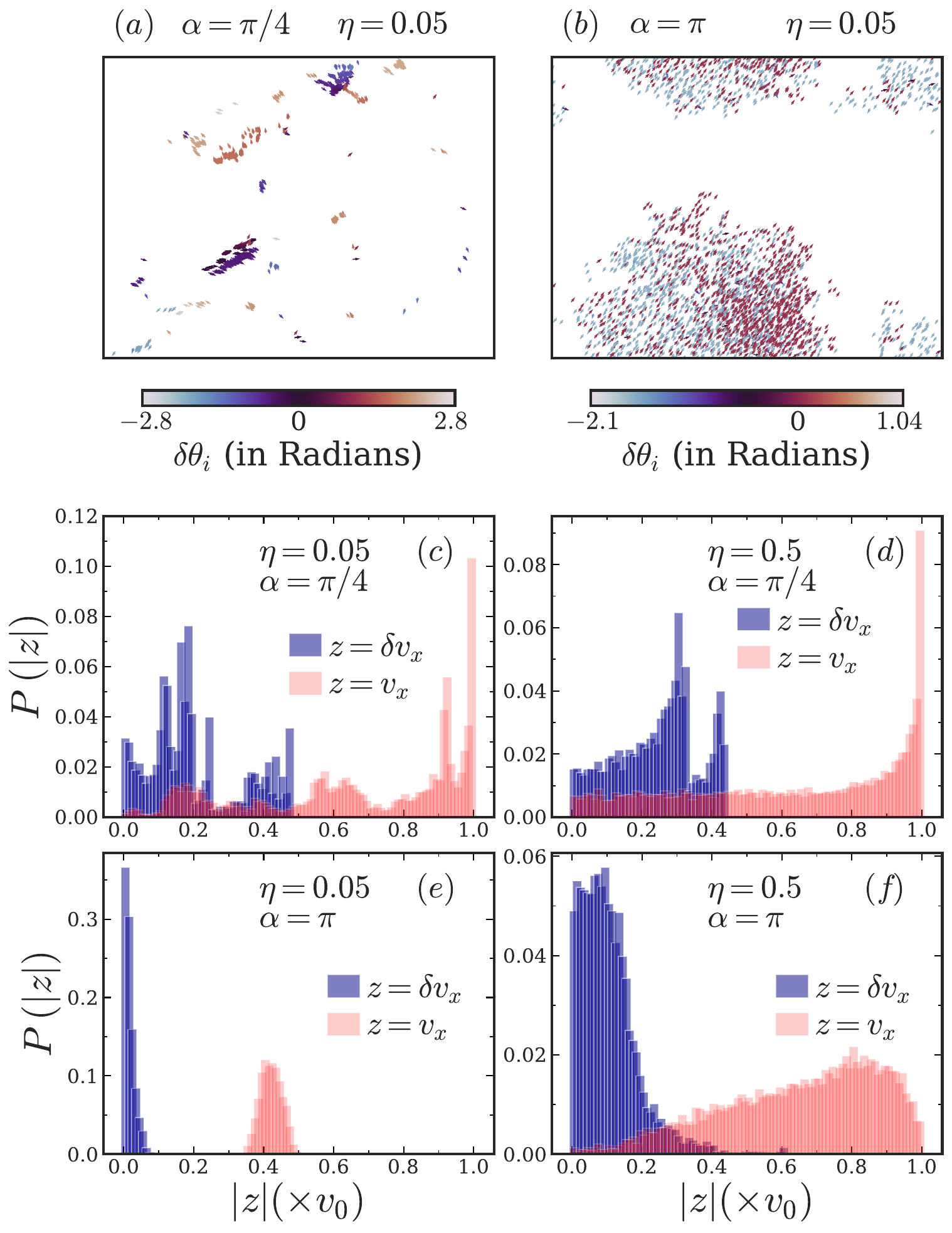}
            \caption{Snapshots showing deviation of the  velocity field $\delta \vec{v}_i$ with  $\eta=0.05$ for $\alpha=\pi/4$ in (a) and $\alpha=\pi$ in (b). For better visualization magnitudes of $|\delta \vec{v}_i|$ are set to $1$.  (c)-(e) Plots of normalized distributions  $P(|z|)$ versus $|z|$ (in units of $v_0=0.01$) with $z$ corresponding to $v_x$ and $\delta v_x$, respectively, for different choices of $\alpha$ and $\eta$, as mentioned. All data are with $N=2560$ and $L=32$.}
            \label{dist_velo_field}
        \end{figure}
        
        \begin{figure}[t]
        %\hskip -0.2in
            \includegraphics[width=9.0cm, height=8.2cm]{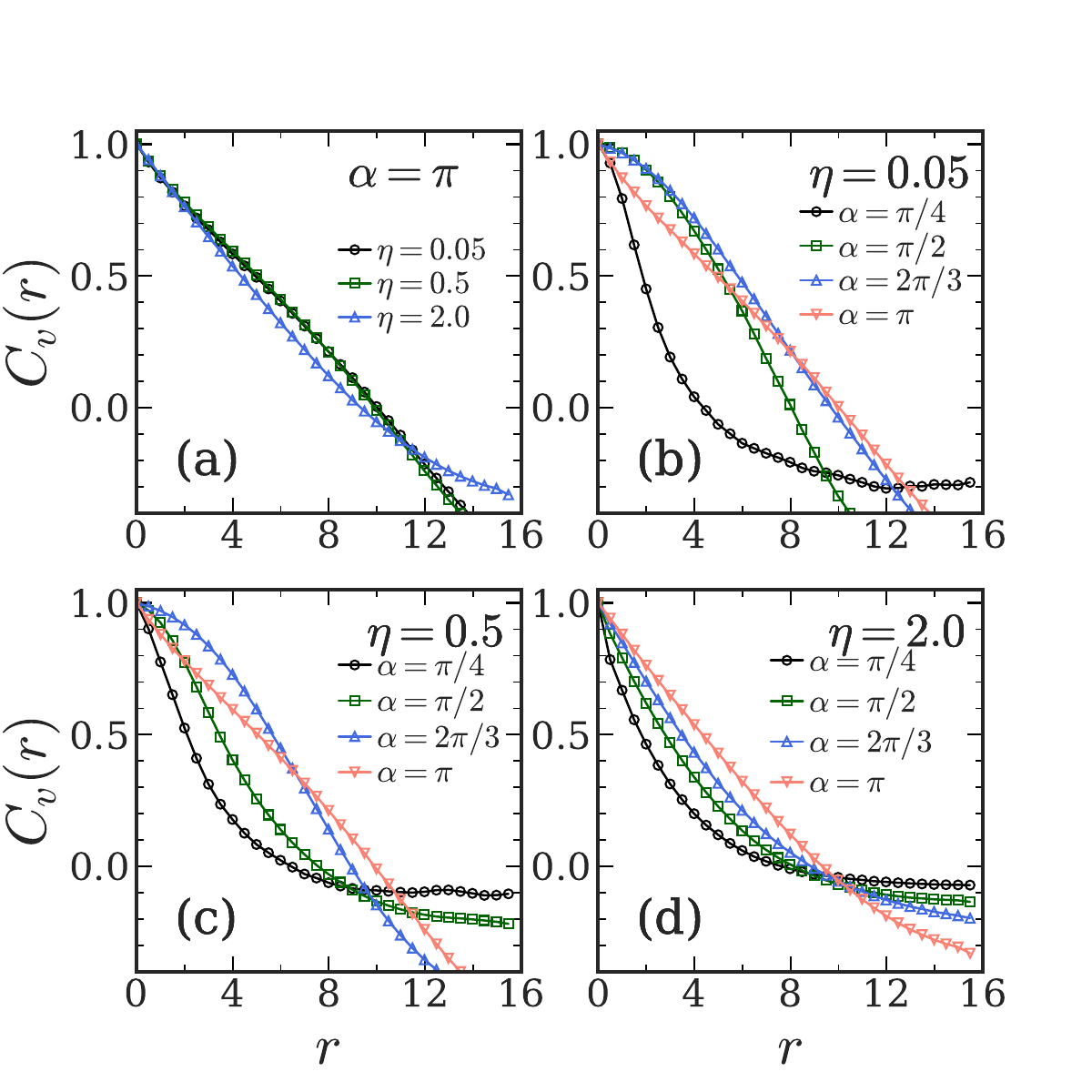}
            \caption{(a) Plots of velocity correlation function   $C_v(r)$ versus $r$ in the steady state for different noise strengths $\eta$, as mentioned, for full angle $\alpha=\pi$. (b)-(d) Plots of $C_v(r)$ for different values of $\eta$ as mentioned. In each frame, comparative plots of  $C_v(r)$  are shown for  different values of $\alpha$. All the presented data are for $N=2560$ and $L=32$. }
            \label{velo_crl_ss}
        \end{figure}

        The degree of alignment of these particles are probed using the velocity order parameter $v_a$ defined as 
        \begin{equation}
           \label{eqn:orderparameter}
            v_{a}=\frac{1}{Nv_{0}}\Big|\sum_{i=1}^{N} \Vec{v}_{i}\Big|\,,
        \end{equation}
        where $|\cdots|$ considers only the magnitude of the velocities. In Fig.~\ref{va_heatmap} we show steady-state phase behavior in the parameter space $\alpha-\eta$ over a wide range ($\alpha \in [0,\pi]$ and $\eta \in [0.01,4.0]$) with respect to the order parameter $\langle v_a \rangle$ which continuously varies between $0$ to $1$ over the parameter space. $\langle \cdots \rangle$ denotes the average over steady-state configurations and  independent runs.
        From this heatmap plot it is identifiable that global order with $\langle v_a \rangle \approx 1$ can be achieved only for low noise and larger $\alpha$ (close to $\pi$), which is also visible from the snapshots presented in Fig.~\ref{snapshot_ss} for $\alpha=\pi$ with $\eta=0.05$ and $0.5$. For intermediate values of $\alpha$ for which clusters of smaller sizes with particles having local velocity ordering is possible, leads to $\langle v_a \rangle \approx 0.5$ (see snapshots corresponding to $\alpha=\pi/2$ or $2\pi/3$). However, for very low $\alpha < \pi/8$ particles remain in a fully random state even with very low noise and these regions are identified with $\langle v_a \rangle$ close to $0$. Hence the global alignment in the velocity space is negligibly small in this regime. However, Ref.~\cite{barberis_2016} suggests that in a similar model system, a locally elongated worm-like cluster with large number of particles may appear in this regime.
        \begin{figure*}[htb]
            \centering
            \includegraphics[width=17cm, height=5.0cm]{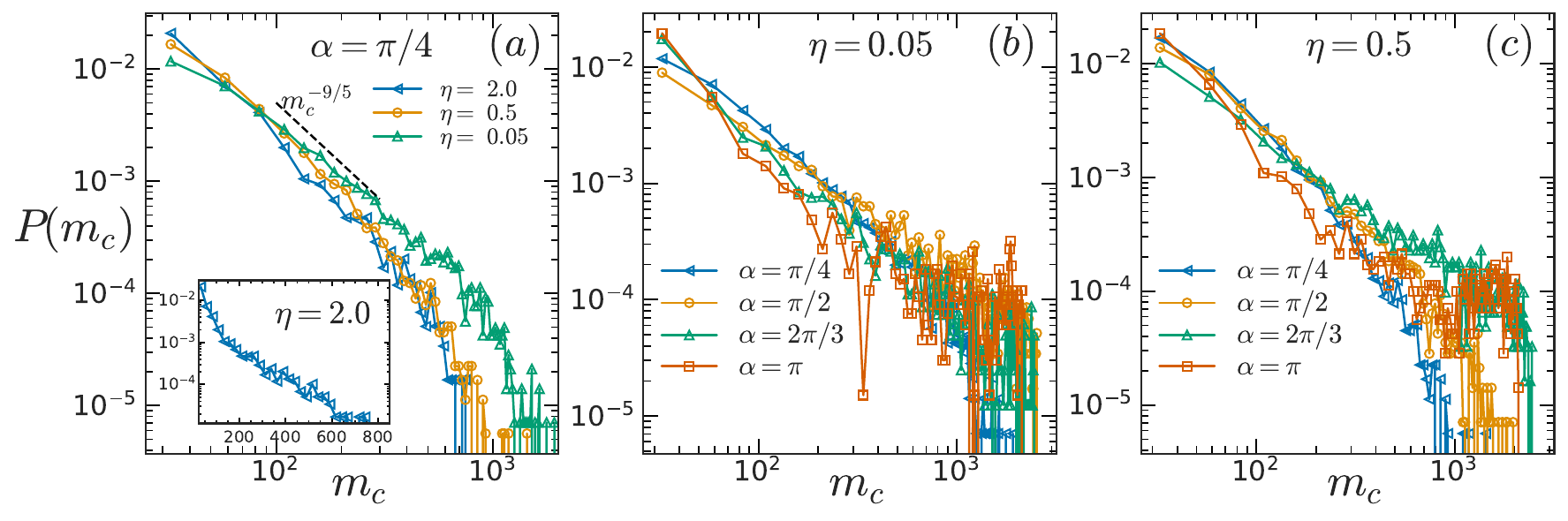}
            \caption{(a) Plots showing the cluster mass distribution $P(m_c)$ versus mass of clusters $m_c$ for $\alpha=\pi/4$ for different values of $\eta$. The dashed line corresponds to a power-law $\sim m_c^{-9/5}$. Inset shows the same data for $\eta=2.0$ on a semi-log scale. Plots of  $P(m_c)$ versus $m_c$ are shown for different values of $\alpha$ as mentioned, for different noise strengths, i.e., for $\eta=0.05$ in (b) and $\eta=0.5$ in (c). All the presented data are for $N=2560$ and $L=32$.}
            \label{dist_clstmass}
        \end{figure*}

        Even though the $\langle v_a \rangle$ elucidates the global ordering that considers velocities of all the particles in the system, local ordering and response can be better identified from the typical flock sizes as well as their local  velocity ordering and related fluctuations. In this regard, we calculate the deviations (velocity fluctuations) $\delta \vec{v}_i$ relative to the global mean. In Fig.~\ref{dist_velo_field}(a) and (b) we show snapshots of the fluctuations corresponding to the velocity field ($\delta v_x, \delta v_y$) for the particles for $\alpha=\pi/4$ and $\pi$ for a lower value of noise $\eta=0.05$. For comparison, we show the corresponding plots of the velocity fields in Appendix~\ref{Appendix_deviation_field}. The color maps in (a) and (b) show the range of corresponding angular deviations ($\delta \theta_i=\tan^{-1}(\delta v_y/\delta v_x)$) in the fluctuation field. As seen, $\delta \theta_i$ varies over a much larger range for $\alpha=\pi/4$ compared to $\alpha=\pi$. For $\alpha=\pi$ the deviations for most of the particles are close to $0$ due to the global coherent motion. For convenience of visualization of the vectors, we set the magnitudes of $|\delta \vec{v}_i|=1$ for all particles. Now, to highlight these qualitative differences, we plot probability distributions $P(|v_x|)$ and $P(|\delta v_x|)$ of the respective variables for a few combinations of $\alpha$ and $\eta$, in Figs.~\ref{dist_velo_field}(c)-(f).  The effects  of lowering the vision angle are prominently observed for lower noises. Qualitatively, the same conclusion can also be drawn from $P(|v_y|)$ and $P(|\delta v_y|)$. Accurate quantification of velocity ordering and related length-scales can be obtained from the velocity and  the associated connected correlation functions, as defined in Eqs.~\eqref{vcf_defn} and \eqref{ccf_defn}, respectively.

The widths of the distributions of $v_x$ and $\delta{v}_x$, as observed from Figs.~\ref{dist_velo_field}(a)-(b), indicate towards the existence of the correlation among the velocities and its fluctuations, which will be quantified here by computing the correlation functions globally.  While the velocity correlation function (VCF) (defined in Eq.~\eqref{vcf_defn}) measures how particle velocities are correlated as a function of their separation distance, the connected correlation function (CCF) (defined in Eq.~\eqref{ccf_defn}) characterizes how the correlation among the velocity fluctuations varies over the separation, thus isolating local ordering from global drift velocity, if any. 

VCF can estimate the typical distance over which direction of the particles are correlated. First, to see the effect of noise,   in Fig.~\ref{velo_crl_ss}(a) we show $C_v(r)$ versus $r$ in the steady state for different values of $\eta$ with $\alpha=\pi$. Plots are normalized by $C_v(r=0) \equiv \langle \vec{v}_i^2\rangle - \langle \vec{v}_i \rangle^2 $ such that $C_v(r=0)=1$. Positive values of $C_v(r)$ correspond to the higher correlation, whereas negative values account for anti-correlation, suggesting almost opposite direction of motion of particles. Such anti-correlation is possible only at larger distances. For all the chosen values of $\eta$, $C_v(r)$s decay with $r$.  The decays are similar to each other. They confirm that for the given density $\rho$ and with $\alpha=\pi$, much higher noise is needed to destroy any flocking behavior, as also seen from Fig.~\ref{va_heatmap}.  

%This confirms that for the lower values of $\alpha$, low noise $\eta=0.05$ is not sufficient enough to restore the velocity ordering during flocking over the length-scale of a cluster. 

%These also confirm that for $\alpha=\pi$ much higher noise is needed to destroy any flocking behavior, as also seen from Fig.~\ref{va_heatmap}. 

The effect of vision-angle $\alpha$ can be understood from figs.~\ref{velo_crl_ss}(b)-(d). In these cases, for each noise strength, comparative plots of $C_v(r)$ vs $r$ are shown for different vision angles $\alpha$. For all the chosen values of $\eta$, decay of $C_v(r)$ becomes faster with decreasing $\alpha$. Note that a faster (slower) decay corresponds to the velocity correlation over a shorter (longer) range, reflecting a smaller (larger) characteristic length-scale associated with the coherently moving particles in a cluster. For high enough noise strength ($\eta=2.0$), the decay seems exponential for all the chosen values of $\alpha$. With $\alpha$ very low ($\alpha=\pi/4$ considered here) such an exponential decay seems to be followed for low values of $\eta$ ($=0.5,0.05$) as well. However, for low $\eta$  the nature of the decay differs qualitatively when the vision angle varies from moderate to maximum (here we consider  $\alpha=\pi/2,2\pi/3$ and $\pi$). In particular, in this parameter space $C_v(r)$ becomes a concave function of $r$, which clearly differs from the exponential one, which is convex in nature. Physically it implies that when the vision angle decreases from its maximum to a moderate value (here we consider $\alpha$ decreasing from $\pi$ to $\pi/2$), even though the cluster size decreases, the order within the velocities of the particles in a single cluster can increase. However, for very small vision angle (here $\alpha=\pi/4$), the cluster size becomes even smaller and the velocity order within the cluster decreases leading towards the exponential decay for all the chosen values of $\eta$. Such an emerging short-ranged higher velocity correlation can also observed from the corresponding plots of $C_{\delta v}(r)$ as well, as presented in Fig.~\ref{ccf-figure_ss} in Appendix~\ref{Appendix_ccf_ss}. However, similarities and differences between those in case of a disordered and globally ordered phase are also discussed.

%Whereas, for lower values of $\eta$ the nature of the decay also changes, for $\eta=2.0$ the decay seems exponential for all values of $\alpha$. Also for $\alpha=\pi/4$, in presence of smaller flocks at lower noises, decay of $C_v(r)$ looks quite different than that for $\eta=2.0$ which corresponds to a state with random particle orientations. A slower decay corresponds to the velocity correlation over a longer range, reflecting a larger characteristic length-scale associated with the coherently moving particles. 

%\textcolor{blue}{Similar behavior is observed for $C_{\delta v}(r)$ as well. This is shown in Appendix~\ref{Appendix_ccf_ss}.}\\ %The presence of higher anti-correlation for lower noises suggests that the different clusters move in different directions, which is also visible from the snapshots presented in Fig.~\ref{snapshot_ss}.\\
   \begin{figure*}[htb]  
          \includegraphics[width=17.5cm, height=7.0cm]{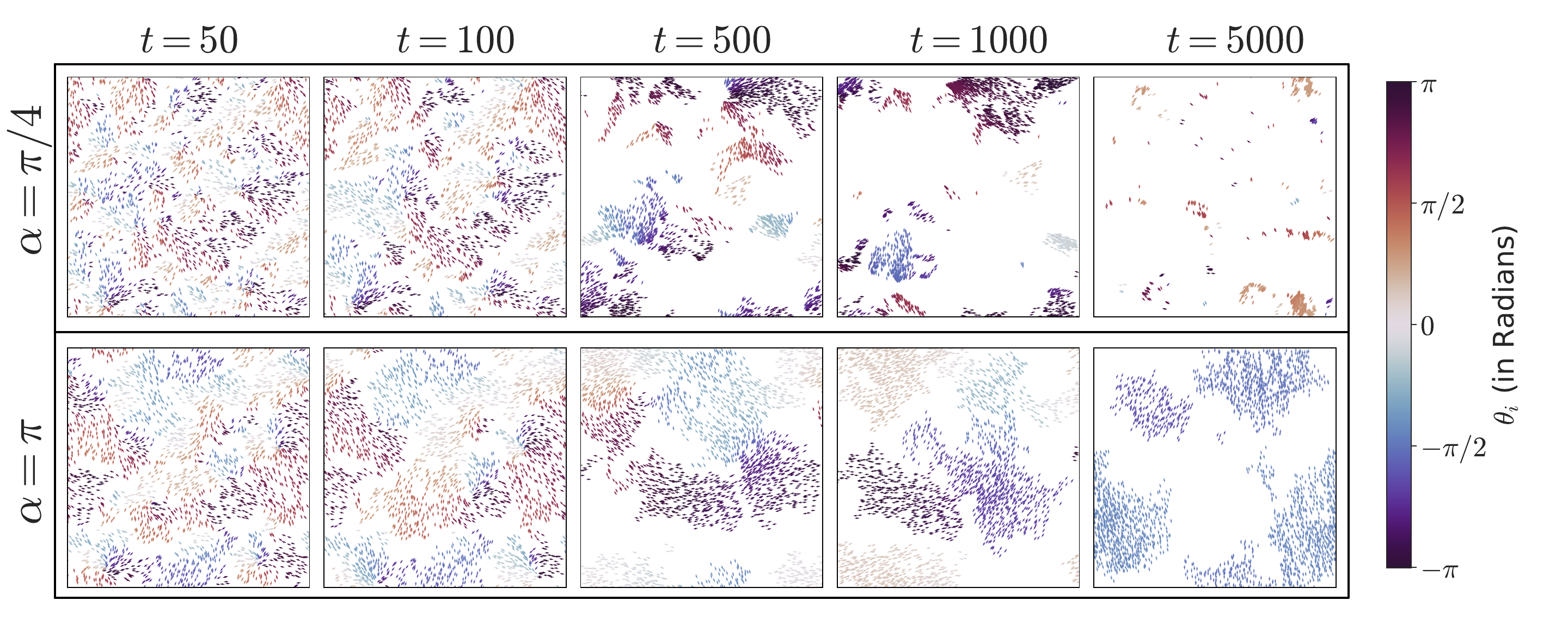}
          \caption{%Typical snapshots of the velocity field of the particles at the steady-state for different $(\alpha=\pi/4,\pi/2,2\pi/3,\pi)$ and $(\eta=0.05,0.5,1)$
          Typical snapshots during time-evolution of the velocity field of the particles  for vision-angle $\alpha=\pi/4$ and $\pi$ for the lower noise $\eta=0.05$. Whereas a global coherent motion appears for $\alpha=\pi$, for lower vision angle $\alpha=\pi/4$ particles form smaller size clusters with local alignment of their velocities. Different colors identify the particles with their respective  orientations. All the snapshots are with  $N=2560$ for a system size $L=32$.}
          \label{snapshot_time}
        \end{figure*}      

Next we focus on clustering of particles in steady state. It is characterized by identifying individual clusters, then finding the number of particles $m_c$  in each cluster and related statistics ($m_c$ also denotes the mass of the corresponding cluster). We calculate the normalized distributions $P(m_c)$ of $m_c$ for different choices of the parameters in the steady-state. In Fig.~\ref{dist_clstmass}(a) we plot $P(m_c)$ versus $m_c$ for $\alpha=\pi/4$ for different noise strengths $\eta$. Even though, the data for $P(m_c)$ for $\eta=2.0$ has a higher value for smaller $m_c$,  after a certain size $m_c \approx 100$,  it always show a lower amplitude compared to the other noises, indicating lower probabilities of formation of larger size clusters. This is also observed in Fig~\ref{snapshot_ss} that, with increasing noise, the emergence of small sized flocks become more prominent. In the inset, data for $\eta=2.0$ is shown in a semi-log scale which identifies the behavior as  exponential-like. Such distribution for $\alpha=\pi/2$ and $\pi$ are shown for different values of $\eta$ in Appendix~\ref{append_mass_alpha}.

For $\eta=0.05$ data looks consistent with a  power-law scaling, i.e., $P(m_c) \sim m_c^{-\mu}$ with $\mu\approx 9/5$ over a larger range compared to the higher noises. This indicates the possibility of formation of larger size clusters for lower noises until finite-size effect appears. Such a power-law scaling corresponds to a scale-free distribution of clusters, suggesting merging of clusters of smaller masses into a larger one.

In Fig.~\ref{dist_clstmass}(b) and (c), the behavior of $P(m_c)$ for different vision angles are explored. For different values of $\alpha$,  $P(m_c)$ versus $m_c$ is shown for $\eta=0.05$ and $0.5$, respectively, in Figs.~(b) and (c). It is evident from the figures that for a fixed noise strength, probability of forming larger clusters decreases (increases) as the vision angle decreases (increases). This fact becomes more evident with increasing noise strength.

        \subsection{Dynamics of flocking and clustering}
        
    Now we focus on the dynamics showing the pathways towards the steady-state behavior  as presented in the previous sub-section. In Fig.~\ref{snapshot_time} we show the time evolution of the system for two  different values of $\alpha$, i.e., $\alpha=\pi/4$ and $\pi$ for $\eta=0.05$, to focus on the effect of $\alpha$ on the emergence of flocking.  Starting from a homogeneous configuration in which the velocities of all the particles are randomly oriented, the coherent structures and the velocity ordering appear and grow with time. Upto $t \approx 100$, the configurations in both cases look somewhat similar. However, at later times, whereas for $\alpha=\pi$ the particles try to form large, globally coherent clusters, for $\alpha=\pi/4$ the clusters appear more localized. The steady-state clusters for $\alpha=\pi/4$ are also smaller in size with higher particle  density and local velocity alignment, quite different than for $\alpha=\pi$. Here low value of  $\eta$ allows us to explore how the spread of the vision cone controls the cluster formation. For high $\eta$ and low $\alpha$, the active particle system does not exhibit any clustering (see Fig.~\ref{time_evolution_high_noise} in Appendix~\ref{Appendix_snap_time} for time evolution with $\eta=2.0$).

        \begin{figure}[htb]
           \includegraphics[width=8.5cm, height=7.2cm]{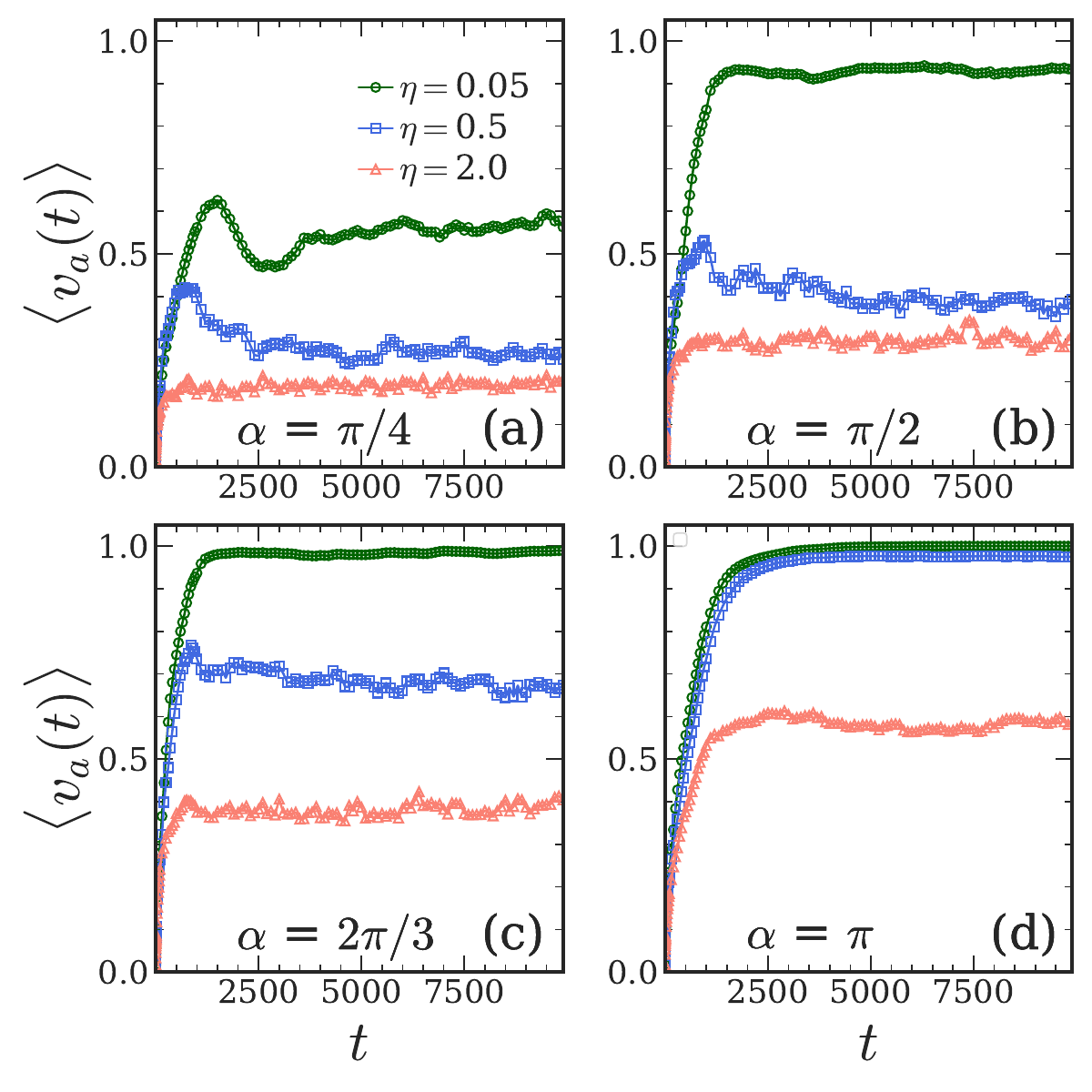}
            \caption{Velocity order parameter $\langle v_{a}\rangle$ versus $t$ for different values of $\alpha$ and $\eta$, as mentioned in (a)-(d). $\langle v_a \rangle \approx 1$ indicates global alignment of velocities of all the particles. All our presented data are for $N=2560$ and $L=32$.}
            \label{orderparafigure}
        \end{figure}

As mentioned earlier, the emergence of the degree of velocity ordering is quantified via the time evolution of the global velocity order parameter defined in Eq.~\eqref{eqn:orderparameter}. In Fig.~\ref{orderparafigure} we show $\langle v_a(t) \rangle$ versus $t$ for different values of $\alpha$ and noise $\eta$. These plots illustrate the competition between the  alignment interaction, the  degree of which depends upon the range of $\alpha$ and $\eta$, to achieve the velocity ordering. For all the cases, the system reaches corresponding steady-state by $t \approx 4000$. For a given $\alpha$, saturation value of $\langle v_a \rangle$ increases with decreasing $\eta$ which can also be inferred from the steady-state data presented in Fig.~\ref{va_heatmap}. For $\alpha=\pi$, with $\eta=2.0$, noise competes and wins against alignment. Hence the global velocity ordering is low and $\langle v_a\rangle$ saturates around $0.5$. One may note here that for $\alpha=\pi/4$ and $\eta=0.05$ also $\langle v_a(t) \rangle$ saturates at $\approx 0.5$. However, the underlying steady state is qualitatively different from the one obtained with $\alpha=\pi$, and $\eta=2.0$. In the steady state with $\alpha=\pi/4$ and $\eta=0.05$,  discrete small  clusters with local (i.e., within a cluster) velocity order have developed. However, globally (i.e., considering all such clusters present in the system) the velocity order looks quite poor. Importantly, in this state, multiple clusters can merge together to form a single larger cluster or a single cluster can break into multiple smaller ones. Both the dynamical processes -- merging and fragmentation -- are spontaneous and seem equally probable. Hence, the state becomes quite dynamic in nature and thus $\langle v_a\rangle$ shows higher fluctuations. The merging and fragmentation events for clusters are better visualized for $\eta=0.5$ and $\alpha=\pi/2$ (see Appendix~\ref{Appendix_snap_alpha_pi2}).  One may ask that how reasonable it is to label such a state as a steady state. However, after a long time, both $\langle v_a\rangle$ and $C_v(r)$ become stable in time, and therefore can be  mentioned as a stable state. In the steady state with $\alpha=\pi$, and $\eta=2.0$, the velocities of the particles are locally and globally disordered and thus $\langle v_a\rangle$ becomes low. In this state probability of obtaining even a single well-defined cluster is quite small. Hence the rate of the dynamical processes like merging and fragmentation of the clusters are also quite low.  
        
        %For this, the discrete clusters with only local velocity ordering within them form. For $\alpha=\pi/4$ with lower noise values interesting intermediate oscillatory behavior appears which can be an outcome related to the intermittent fragmentation and merging of clusters which will be clearer later.
        % For $\alpha=\pi$ the magnitude of $\langle v_a\rangle$ is steady for 

\begin{figure}[t]
        \centering
            \includegraphics[width=8.8cm, height=7cm]{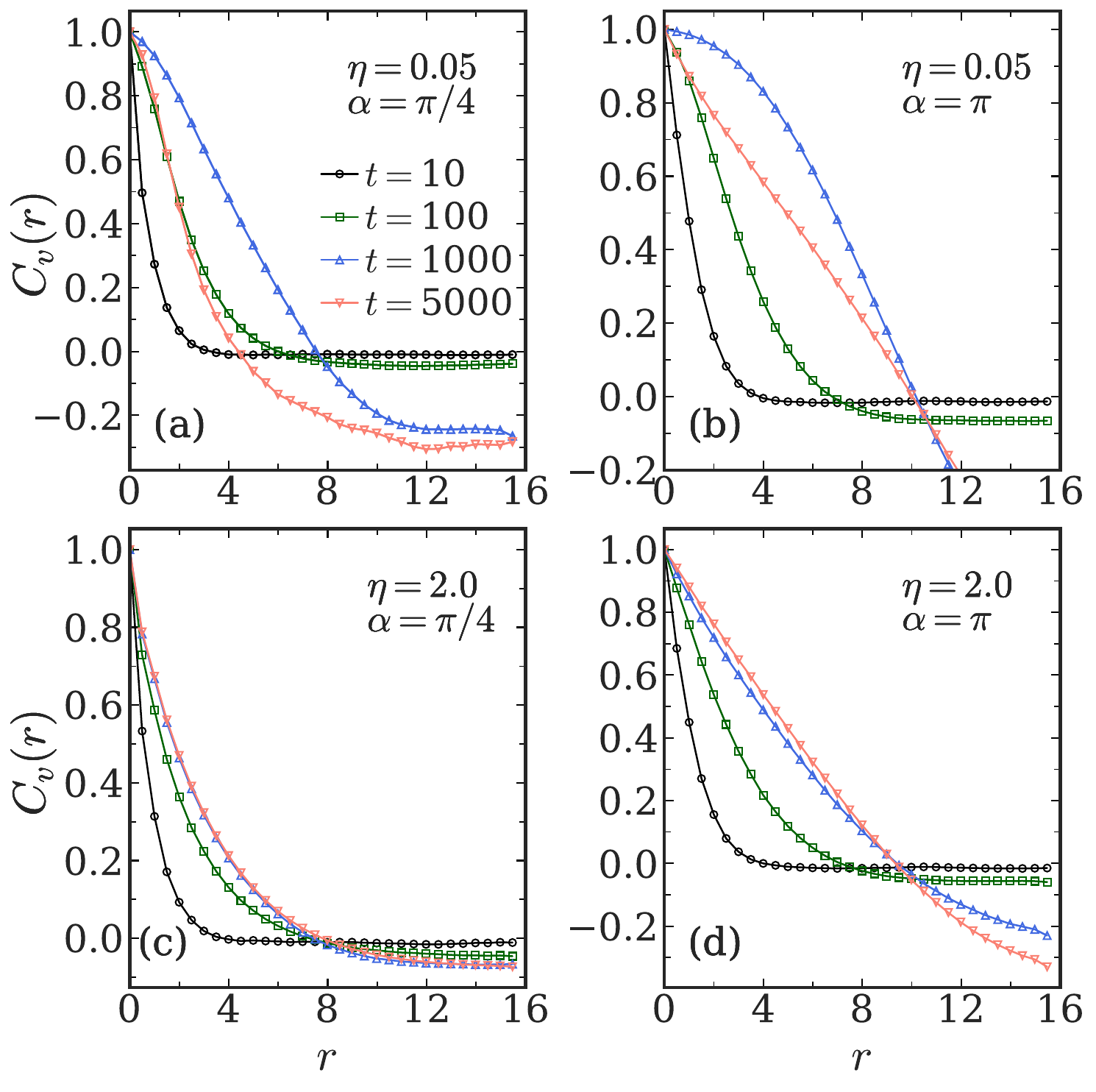}
            \caption{(a)-(d) Time evolution of  $C_v(r,t)$ as a function of $r$, for $\alpha=\pi/4$ and $\pi$ for lower and higher noises, i.e., $\eta=0.05$ and $2.0$, respectively. Corresponding times are mentioned. All data are for $L=32$ with $N=2560$. }
            \label{velo_crl_time}
        \end{figure}

Next we discuss the time evolution of velocity correlations.  To see how the correlation develops over time, we present results for the time evolution of $C_v(r,t)$ versus $r$ in Fig.~\ref{velo_crl_time}(a)-(d) for two noises ($\eta=0.05$ and $2.0$) for $\alpha=\pi/4$ and $\pi$. The slower decay of VCF with increasing time suggests the development of a velocity correlation among particles over larger distances. For $\eta=2.0$, the decay of VCF at late times is much slower in the case of $\alpha=\pi$ compared to that of $\alpha=\pi/4$. This indicates that for low $\alpha$, the clusters are smaller and the correlations among themselves are quite low compared to   high $\alpha$.%For the higher noise with $\eta=2.0$ whereas this fact is observed  for $\alpha=\pi$, data for $\alpha=\pi/4$ at late times more or less overlap on each other.

However, with lower noise $\eta=0.05$ a non-monotonic behavior is observed over time  in the decay of $C_v(r,t)$ for both values of $\alpha$. For these, at very late times, at $t \ge 5000$, decay of $C_v(r,t)$ becomes faster compared to intermediate times (e.g. $t=1000$), indicating a lowering of the correlation length-scale at  late times. The reduction of the average length-scale over which particle velocities are correlated can be possible either by breaking of the larger clusters into smaller ones or by shrinking of the individual clusters with the same number of particles which eventually increases the local density. The latter is also possible in this model as the particles do not have an excluded volume.  Similar non-monotonic behavior is observed for the time-dependent CCF as well, which are presented in Fig.~\ref{ccf-figure} in Appendix~\ref{Appendix_ccf}.
        
To verify this, we calculated the time evolution of the average number of clusters $\langle n_c \rangle$ present in the system. Here, $\langle \cdots \rangle$ corresponds to an average over independent runs. As mentioned, a cluster is identified when the number of particles, i.e., its mass $m_c \ge n_{\text{crit}}$. A detailed description on cluster identification is provided in Appendix~\ref{Appendix_clust_id}. In Fig.~\ref{avg-cluster-number}(a)-(b) we show $\langle n_c(t) \rangle$ versus $t$ for $\alpha=\pi/4$ and $\pi$, for different values of $\eta$. First, we discuss the case for $\alpha=\pi$. For this, for all values of $\eta$, the number of clusters $\langle n_c(t) \rangle$ increases to a maximum $n_c^{\text{max}}$ and then decreases. Such trend is also observed in general for a system of particles with attractive interactions \cite{roy_droplet_sm,paul2021clusters} or in case of collapse of a polymer during its coil-globule transition \cite{paulsm_2022vic}. Decrease in the numbers suggests the merging of small clusters to larger ones. At later times, whereas for lower noises such merging of clusters is faster,  for $\eta=2.0$ $\langle n_c(t) \rangle$ decreases much slowly, indicating that the cluster-merging events become less probable for higher noise strengths. Also, $\langle n_c(t) \rangle$ seems to saturate at $n_c^{s} \approx 8$. Values of $n_c^{\text{max}}$ and $n_c^{s}$ are restricted by the finite size of the system. 

For $\alpha=\pi/4$, while a similar trend is observed for lower noises $\eta=0.05, 0.5$, except the fact that at late times $\langle n_c(t)\rangle$ shows a very slow increase in these cases. Increase in $\langle n_c(t)\rangle$ at late times suggests fragmentation of clusters. At late times, in this parameter space, even though the clusters can break, low noise does not affect their global orientational ordering much, resulting in a saturation of $\langle v_a \rangle$, as seen from Fig.~\ref{orderparafigure}(a). However, at intermediate times both fragmentation and coalescence of clusters affect the time evolution of $\langle v_a(t)\rangle$ strongly.  At higher noise $\eta=2.0$ lower vision angle prevents the development of stable clusters as well as the formation of any coherent flocks. In this case $\langle n_c(t)\rangle$ grows initially but saturates to a very small value.

%Not reaching a maximum for $\langle n_c(t)\rangle$ indicating no formation of fully stable clusters in the system. 
            
        \begin{figure}
            \includegraphics[width=8.2cm, height=7.8cm]{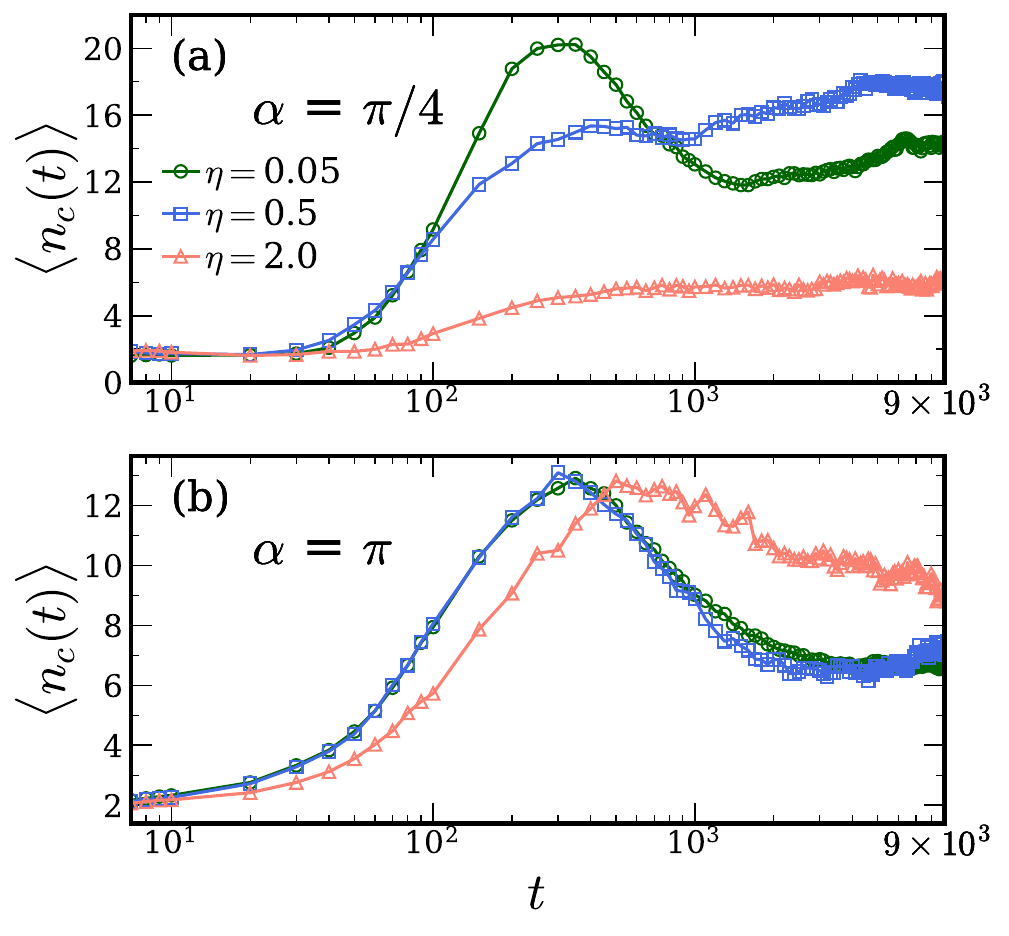}
            \caption{Plots show time evolution of the number of clusters $\langle n_c(t)\rangle$ for $\alpha=\pi/4$ in (a) and for $\alpha=\pi$ in (b) for different noise strengths $\eta$. All data are for $N=2560$ and $L=32$.}
            \label{avg-cluster-number}
        \end{figure}

\begin{figure}[h]
        
            \includegraphics[width=8.2cm, height=7.4cm]{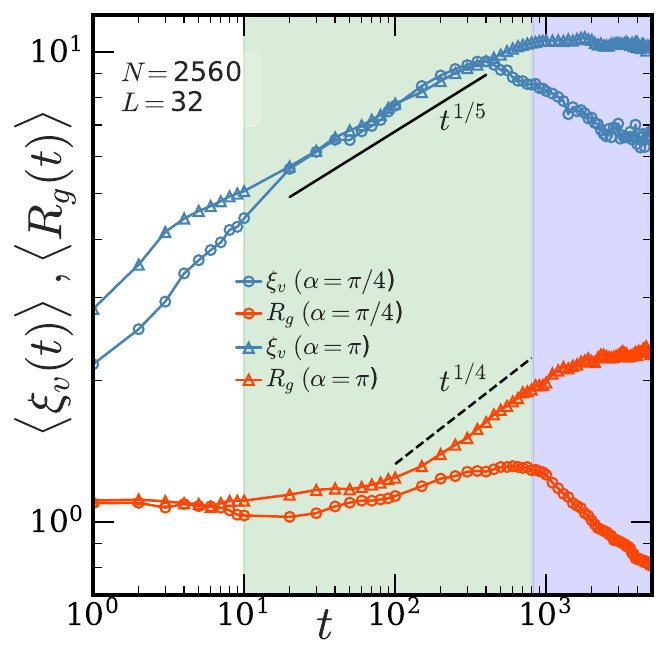}
            \caption{Plots of correlation length $\langle \xi_{v}\rangle$ calculated from the decay of $C_v(r,t)$ and  $\langle R_g \rangle$ versus $t$ for  $\alpha=\pi/4$ and $\pi$ with $\eta=0.5$. Shades are used as guide to the eye to distinguish between the scaling and the finite-size affected regimes. All data are for $N=2560$ and $L=32$. }
            \label{rg_correlation_length}
        \end{figure}

To further quantify, we calculate the correlation length $\xi_v$ as well as the radius of gyration $R_g$ for the clusters.  Whereas $\xi_v$ quantifies the span of velocity ordering of the particles, $R_g$ measures the typical length-scale or spatial size of a cluster. In this case, as the particles do not have any size,  the overlap of particles results in very high density condensates and $R_g$ may not yield an accurate estimate of the cluster size particularly at  lower noise strengths. However, for higher noise cases the estimate works better. $\xi_v(t)$ is calculated from the decay of the corresponding $C_v(r,t)$ at time $t$ when it decays to the $1/e$ fraction from its starting value, i.e., 
        \begin{equation}
            C_v(r=\xi_v)=\frac{1}{e}~C_v(r=0)\,,
        \end{equation}
where $C_v(r=0)=1$ due to the normalization. $R_g$ is calculated using the definition in Eq.~\eqref{radius_of_gyration} for individual clusters present in the system at the corresponding times. In Fig.~\ref{rg_correlation_length} data for time evolution of $\langle \xi_v(t)\rangle$ and $\langle R_g(t) \rangle$ are presented for the intermediate noise $\eta=0.5$ for $\alpha=\pi/4$ and $\pi$. Where $\langle \cdots \rangle$ in $\xi_v$ corresponds to the average over  independent runs, such an average for $R_g$ is done for different clusters present at any instant and over different independent runs. For both plots, three distinct regimes can be seen. In the first regime velocity correlation and clustering start developing. The development of velocity correlation appears to be earlier than the corresponding density field clustering. This fact looks somewhat similar to that for a granular gas, where dissipation in energy leads to parallelization of velocities, which eventually leads to clustering in the density field \cite{paul2014dynamics,paul2017ballistic}. Also, for both values of $\alpha$, the amplitudes corresponding to $\xi_v$ are always higher than those for $R_g$.
        
        After the initial transient regime, the system enters the scaling regime (shown by green shaded region as a guide to the eye). Data for $\xi_v$ in the scaling regime follow $\langle \xi_v \rangle \sim t^{1/5}$.  Whereas $\langle \xi_v \rangle$ data for both values of $\alpha$ fall on each other, for $\langle R_g \rangle$, data for $\alpha=\pi/4$ has values lower than that for $\alpha=\pi$. This indicates to a lack of formation of a large and stable cluster for $\alpha=\pi/4$.  For $\alpha=\pi$, in the scaling regime, it shows $\langle R_g(t) \rangle \sim t^{1/4}$, an exponent slightly larger than that for $\langle R_g \rangle$. For $\alpha=\pi$, data for both $\xi_v$ and $R_g$ finally saturates due to the finite-size effect.  However, for $\alpha=\pi/4$, both data for both $\langle \xi_v \rangle$ and $\langle R_g \rangle$ decay at late times after $t \approx 10^3$ due to the corresponding fragmentation events which was already hinted from the plots of Fig.~\ref{avg-cluster-number}(a) and observed from the movies in supplemental material. Similar to $\xi_v$, length-scale $\xi_{\delta v}$ can be calculated from $C_{\delta v}(r,t)$ and provides estimate on the distance over which the fluctuations are correlated. A comparison of the time evolution of $\langle \xi_v(t)\rangle$ and $\langle \xi_{\delta v}(t)\rangle$ is provided in Fig.~\ref{connectedcorrlength_time} in Appendix~\ref{Appendix_correl_length}. For both parameters, higher values of $\xi_v$ in comparison to $R_g$ suggests that velocities may be correlated beyond clusters of typical size and correlation over larger distance for $\alpha=\pi$ than for $\alpha=\pi/4$. 

        \section{Conclusion and Outlook}
        
        Using intensive numerical simulations we have studied the dynamics related to clustering and flocking of active particles due to velocity alignment in presence of noise and non-reciprocal interactions implemented via a  mechanism of visual perception. Our results show that the strength of the noise $\eta$ and the vision angle $\alpha$ plays very crucial and competing roles in determining the degree of alignment interaction and eventually the formation and stability of clusters. For a full vision angle $\alpha=\pi$ and low noise, the system exhibits global ordering with all the particles moving coherently, as observed in the original Vicsek model \cite{vicsek-1995}.  However, with lowering of $\alpha$ the global ordering decreases and the system tends to form small clusters with local alignment of velocities. As $\alpha$ decreases, the stability of the larger clusters   diminishes significantly, also leading to intermittent fragmentation. For high noise and small $\alpha$, the system remains in the homogeneous disordered state. Lowering of the vision angle,  i.e., with increasing degree of non-reciprocity in the interaction, hinders the ability of the particles to flock in a globally ordered manner, as evidenced by the decrease in the global order parameter $\langle v_a\rangle$. However, for lower noises, even though the clusters are locally ordered, the particles within a cluster seem to maintain a higher correlation among their velocities in comparison for a globally ordered state appearing in case of full vision angle. 
        
        Our steady-state phase diagram over a wide range in the $\alpha$-$\eta$ parameter space highlights the transition between various phases: from a disordered phase to locally ordered clusters to a globally ordered state. The distributions of the velocity components and corresponding fluctuations  relative to the global mean, for different values of $\alpha$, helps to distinguish between the global and local velocity ordering. Pathways towards their emergence have been  quantified from time evolution of the VCF and CCF, i.e., $C_v(r,t)$ and $C_{\delta v}(r,t)$, respectively. Furthermore, analysis of the correlation length $\xi_v(t)$ calculated from the decay of $C_v(r,t)$ and the average radius of gyration $R_g(t)$ along with the time evolution of the number of clusters in the system, provides important evidence regarding the stability and fragmentation events in the limits of lower vision angles. This also confirms towards the emergence of particle clustering as a consequence of the developed correlation in the velocity field. However, it can be interesting to investigate the dynamics of the clusters, information of which may shed light on the cluster growth and fragmentation mechanisms.%While the cluster mass distribution maintains a scale-free nature, consistent with previous findings, the radius of gyration, representing cluster size, was consistently smaller than the velocity correlation length, attributed to the zero-size particle assumption in our model.\\

In an earlier work \cite{zumaya2018delay}, it was shown that in the absence of periodic boundary condition, i.e., when particles move in free space, the coherent motion can be destroyed even by some small amount of noise. It has also been proposed that allowing particles to interact with a few neighbors at larger distances can improve cohesiveness of the flock and promotes long-range correlations. However, it is not clear how such neighbors can be chosen in a systematic and efficient manner. One possibility is to select a few neighbors randomly at a larger distance which may mimic the effect of a leader in a flock. In another approach,  interactions can be made with a few fixed neighbors based on their topological distance rather than a metric distance \cite{Ballerini_2008_birds_collectivemotion}. Thus in our model which incorporates visual perception, it would be interesting to examine how tuning the interaction in different ways influences the stability and coherence of flocks. It would also be worthwhile to see whether introducing a finite delay in response during velocity alignment with neighbors within vision cone could enhance the stability of clusters and the flocking behavior \cite{zumaya2018delay}. Furthermore, in case of asymmetric, non-reciprocal interactions at lower vision angles, it would be interesting to see whether the locally ordered structures that emerge at low noise can display any emergent chirality, i.e., a non-zero persistent rotational motion, similar to that observed in many biological flocks across a wide range of length scales.

%Another interesting aspect will be to check the effect of a vector noise on the emergent patterns in the presence of visual perception. Vector noise, as for the original Vicsek model, can change the collective behavior towards formation of bands or traveling waves and also make it as a first order transition in stead of a continuous transition for scalar noise.

 In this work, like in many similar studies, we did not consider any effect due to the volume exclusion of the particles, which seems important as any living object has a well-defined size. Such an effect can be included using a soft-core repulsive potential. In such a scenario, it can be possible to relate the behavior of flocking and clustering with the well-known coarsening processes and estimate the corresponding dynamic scaling exponents \cite{katyal2020coarsening,chate2024dynamic}, perform a finite-size-scaling as well\cite{fss_das2024} and compare with a few known results in similar systems to achieve further insight regarding the dynamics and growth mechanism. Also it can be interesting to investigate the relaxation kinetics and calculate properties related to aging during flocking starting with different initial conditions, similar to phase transition kinetics with Ising model \cite{das2020initial}.   While the interaction implemented via the vision-cone with $\alpha \ne \pi$ captures the effect of non-reciprocity, it is not quantified by any measure relative to the reciprocal interaction. As a future work, we plan to quantify this as a variable using a graph theoretic approach. In particular, it can be done by constructing directed networks from the adjacency matrix and then by calculating the corresponding trophic levels of the graph \cite{mackay2020directed}. Such a method can be useful to quantify the degree of non-reciprocity in general in similar interacting many-body systems.  %In a few works, it has been shown that predicting neighbors in a more efficient can provide better stability of flocks, e.g., way rather than by interacting only based upon the metric distance is more useful, interacting with a few neighbors at long distance, etc. It can be worth to check such considerations in presence of vision cone which can provide a better framework to understand dynamics of the flocks and mimic the collective pattern shown by many living objects. 

  %  Also analysis related to the local density and angle distributions can provide better understanding regarding the pattern and structure of the clusters. 
        
      \section{Acknowledgements}
      S.P. and A.S. thank Debasish Chaudhuri for discussions.  S.P. acknowledges the University of Delhi for providing financial assistance through the Faculty Research Programme under Grant-IOE (Ref.\ No.\ IOE/2024-25/12/FRP). A.S. acknowledges Department of Science and Technology, Govt. of India for the ARG-ANRF grant (file number: ANRF/ARG/2025/009343/PS). A.S. also acknowledges the funding ({\it{Investissements d’Avenir}}, ANR-16-IDEX-0008) by CY Initiative of Excellence. This work was partially developed during his stay at the CY Advanced Studies, whose support is acknowledged.

\section{Conflicts of Interest}
There is no conflicts of interest to declare.

\section{Data Availability}
Data are available upon request to the authors.
        \appendix

        \section{Velocity and Corresponding Fluctuation Fields}\label{Appendix_deviation_field}
\begin{figure}[h]
        	\includegraphics[width=7cm, height=8.5cm]{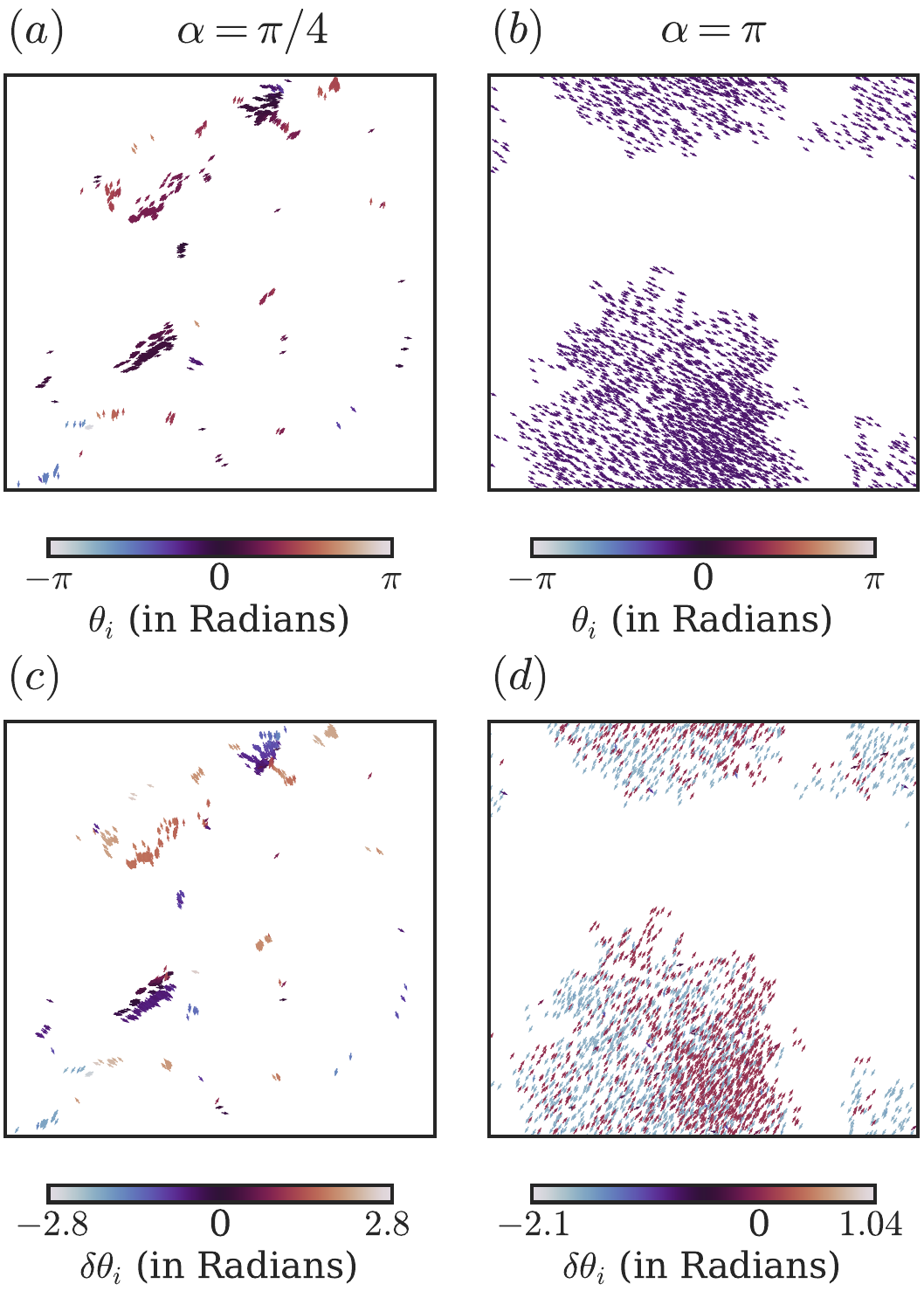}
        	\caption{Snapshots of the (a)-(b) velocity field $\vec{v}_i$ (with orientations $\theta_i$) and (c)-(d) velocity fluctuation field $\delta {\vec{v}_i}$ (with angular deviations $\delta \theta_i$)  for $\eta=0.05$ and $\alpha=\pi/4$ and $\pi$. All data are for $L=32$ and $N=2560$.}
        	\label{fig_deviation_field}
        \end{figure}

        As a reference to Fig.~\ref{dist_velo_field}, here in Fig.~\ref{fig_deviation_field} we show comparative plots of  velocities  $\vec{v}_i$ for $\alpha=\pi/4$ and  $\pi$ in (a) and (b), respectively,   as well as their corresponding fluctuations $\delta \vec{v}_i$ in (c) and (d). Due to presence of global order and correlation of velocities over a larger distance for $\alpha=\pi$, corresponding spreads for the angular deviations are smaller and likely close to $0$ compared to those  for $\alpha=\pi/4$ (as seen from the respective color bars). 
            
         \section{Steady-state behavior of Connected Correlation function (CCF)} \label{Appendix_ccf_ss}
         \begin{figure}[htp]
            \includegraphics[width=8cm, height=7cm]{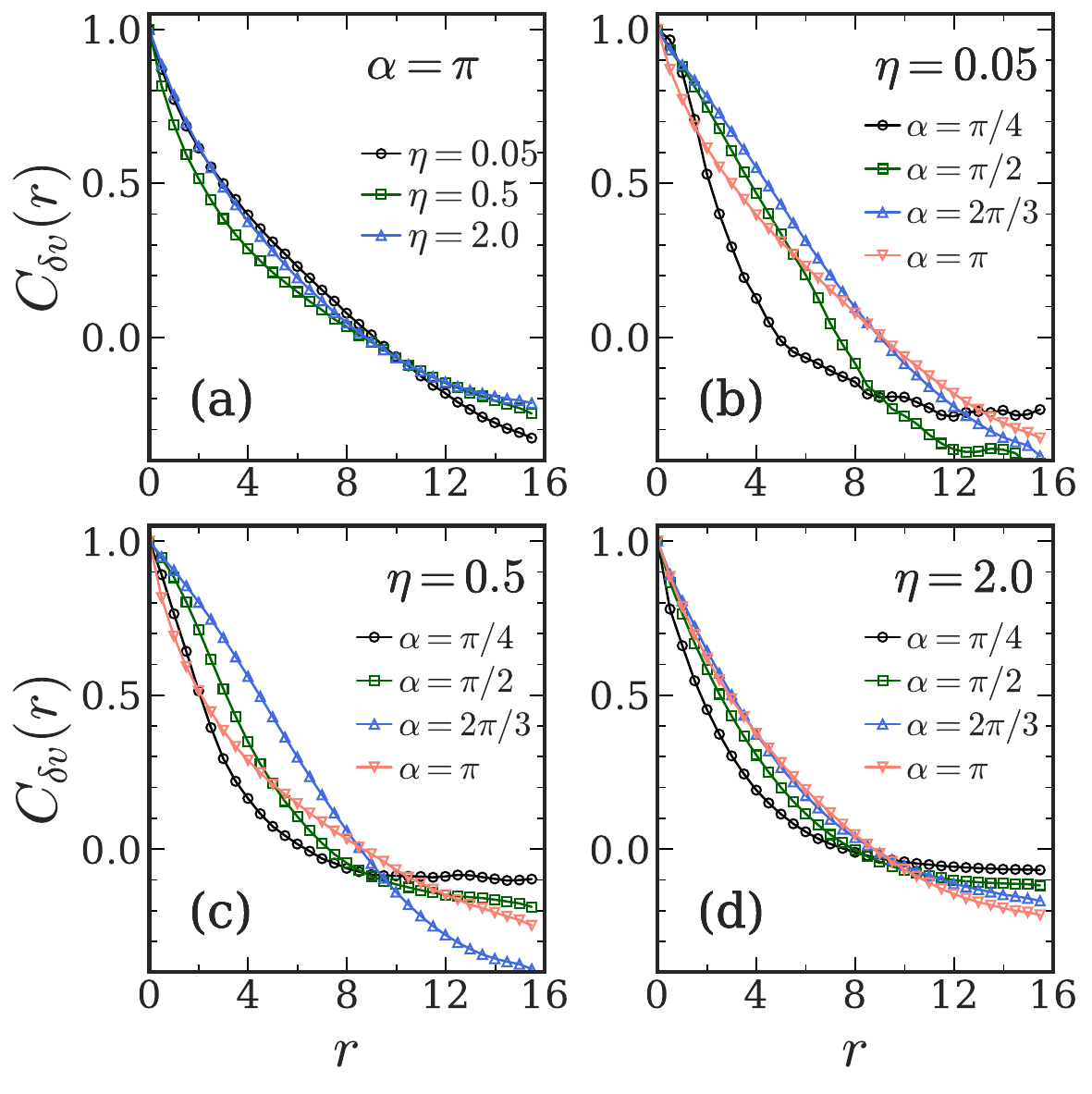}
            \caption{ (a) Plots of CCF   $C_{\delta v}(r)$ versus $r$ in the steady state for different noise strengths $\eta$, as mentioned, for $\alpha=\pi$. (b)-(d) Plots of $C_{\delta v}(r)$ for different values of $\eta$, as mentioned. In each frame, comparative plots of  $C_{\delta v}(r)$  are shown for different values of $\alpha$. All data are for $L=32$ and $N=2560$. }
            \label{ccf-figure_ss}
        \end{figure}
        The connected correlation function (CCF) quantifies the correlation between the fluctuations in the velocity field as defined in Eq.~\eqref{ccf_defn}. This is useful to distinguish the collective coherent behavior and the local orderings. To separate the fluctuations from the common global average part, we calculate the CCF which shows how the fluctuations are correlated in the system. In Fig.~\ref{ccf-figure_ss} we show $C_{\delta v}(r)$ versus $r$ at the steady-state for different set of parameters. For uncorrelated velocities with $\langle v_a \rangle \approx 0$ one can have  $C_v(r) \approx C_{\delta v}(r)$. However, for ordered states with $\langle v_a \rangle \ne 0$, $C_{\delta v}(r)$ can have different behavior than that from corresponding $C_v(r)$. In fact, for a completely correlated velocities, i.e., in presence of a global ordering whereas VCF shows a much slower decay, CCF can decay faster (as seen from Fig.~\ref{velo_crl_ss} and \ref{ccf-figure_ss} for $\eta=0.05$ and $\alpha=\pi$). In particular, for living systems, length-scale calculated from CCF is more relevant for characterizing the measure of coherence and information propagation, since it removes the contribution from global alignment and captures genuine interaction-mediated correlations. Comparison between $C_v(r)$ and $C_{\delta v}(r)$ show that even in presence of power-law-like decay of VCF corresponds to a faster decay of $C_{\delta v}(r)$.

\section{Cluster Identification }\label{Appendix_clust_id}
        
\begin{figure}[htb]
            \includegraphics[width=0.85\linewidth]{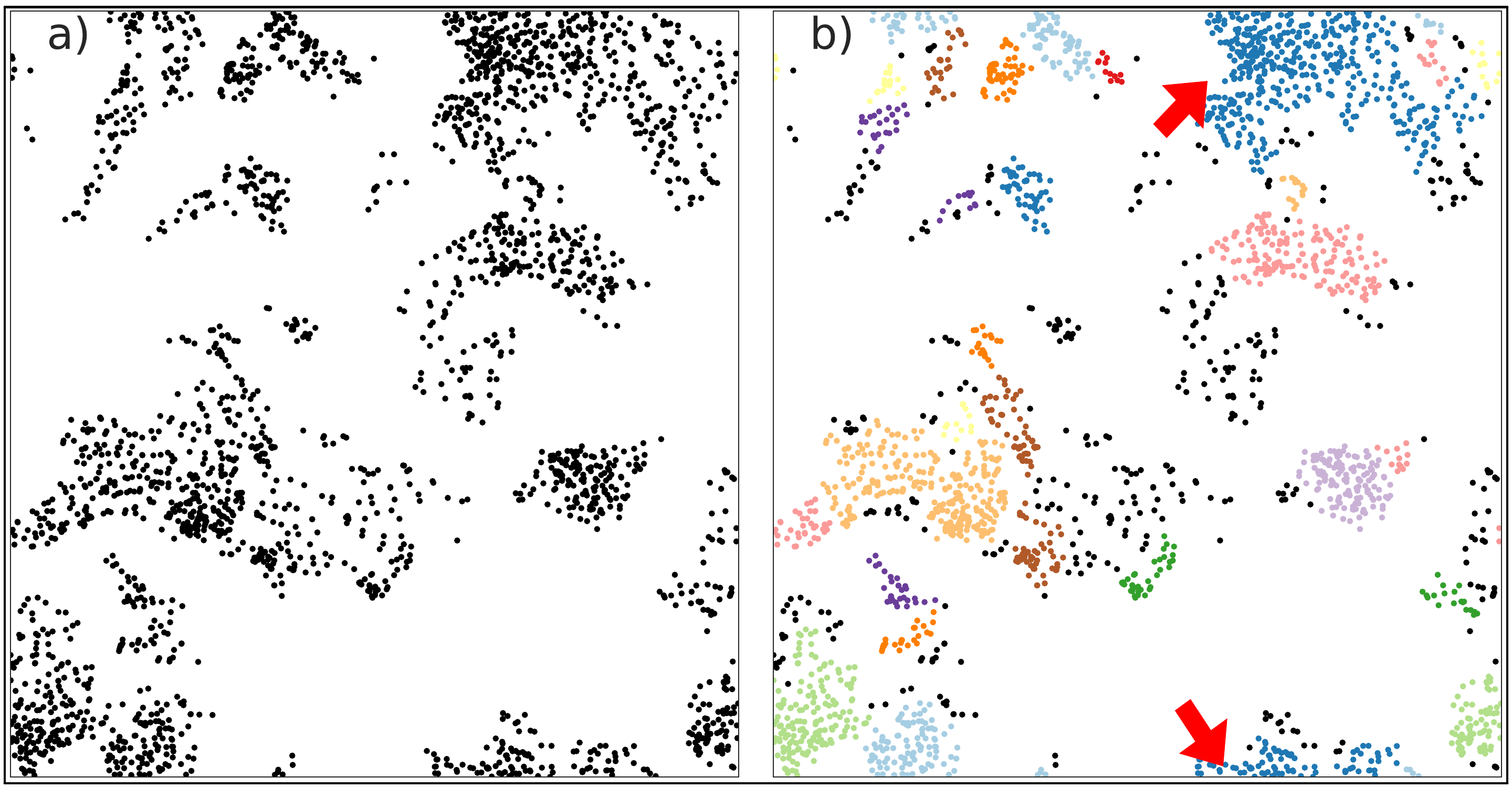}
            \caption{(a) A typical snapshot illustrating the density field clustering for $\eta=0.05$ and $\alpha=\pi/4$. (b) Different colors  highlight different clusters and the particles which are not part of any cluster are marked as black. }
            \label{cluster_identification}
        \end{figure}

        As mentioned, the cluster identification was done using breadth-first-Search (BFS) algorithm \cite{ismail1989multidimensional}. Using the cutoff distance $r_{\text{cut}}=0.5$ an adjacency matrix or directed  graph is created with $1$ signifying a connection and $0$ as no connection between any two  $i$-th and $j$-th nodes in the graph. Then, the BFS algorithm systematically traverses through the graph and identifies all the connected nodes using the adjacency matrix. If the connected graph has nodes together more than $n_{\text{crit}}$, then they are considered as a cluster, as shown in Fig.~\ref{cluster_identification}. These data are subsequently used to calculate the  number of clusters $\langle n_c \rangle$ as well as their average radius of gyration $\langle R_g \rangle$.  In Fig.~\ref{cluster_identification}(b) different clusters are identified and marked by different colors. As shown, among those a few clusters span on the other side of the system due to the application of PBC. Whereas calculation of the center-of-mass(CM) of a  cluster localized on one side is straightforward, evaluation of CM for any spanning cluster requires careful analysis to properly incorporate the effect of periodic boundary condition (PBC). Calculation of CM for such a cluster (marked by red arrows in Fig.~\ref{cluster_identification}(b)) is described in Appendix~\ref{Appendix_cm}.
        
        %one of the cluster spanning the periodicity of the system is highlighted with an arrow and blue colour of the particles. Such clusters can lead to misleading calculation of the center of mass (CM) if the PBC are not taken into account. Thus while calculation of the CM we have used circular mean, which is discussed below, to take care of the PBC .

        \section{Calculation of center-of-mass of a cluster under PBC }\label{Appendix_cm}
        
        A simple arithmetic mean of the particle coordinates will lead to wrong value for the CM of the clusters which is spanned in both sides of the system due to PBC. To correctly compute the CM, we use the circular mean method, which projects the linear coordinate components onto a unit circle \cite{bai2008calculating}. This method is applied independently for the coordiantes in both the $x$ and $y$ directions. Below we discuss the method.
        
        For the position $u_{i,\alpha}$ of the $i$-th particle where $\alpha \in \{x, y\}$, is transformed into two coordinates defined on a unit circle $\mathcal{C}$ as
        
        \begin{equation}
            \gamma_{i, \alpha} = \cos\left(\frac{2\pi u_{i, \alpha}}{L}\right)~~~~\text{and}~~~~
            \zeta_{i, \alpha} = \sin\left(\frac{2\pi u_{i, \alpha}}{L}\right)\,,
        \end{equation}
        where $L$ corresponds to the length of the box.
        
        Then the average of these transformed coordinates $\gamma$ and $\zeta$ are calculated for all the particles in the cluster, say $N_c$,  over $\mathcal {C}$ as
        
        \begin{equation}
            \bar{\gamma}_{\alpha} = \frac{1}{N} \sum_{i \in \mathcal{C}} \gamma_{i, \alpha}\,,\;\;\;\;\;\;\; 
            \bar{\zeta}= \frac{1}{N} \sum_{i \in \mathcal{C}} \zeta_{i, \alpha}\,.
        \end{equation}
        
        Finally, the $x$ and $y$ components of the CM $r_{\text{cm}, \alpha}$ is obtained by reversing the transformation using
        \begin{equation}
                    r_{\text{cm}, \alpha} = \frac{L}{2\pi} \tan^{-1}\left( \frac{\bar{\zeta}_{\alpha}}{\bar{\gamma}_{\alpha}}\right) \,.
                            \end{equation}
This is known as a circular mean, and it takes care of the PBC for a spanning cluster and calculates its $R_g$ correctly. 
        
\section{Cluster mass distribution}\label{append_mass_alpha}
        \begin{figure}[h]
            \includegraphics[width=8.20cm, height=4.0cm]{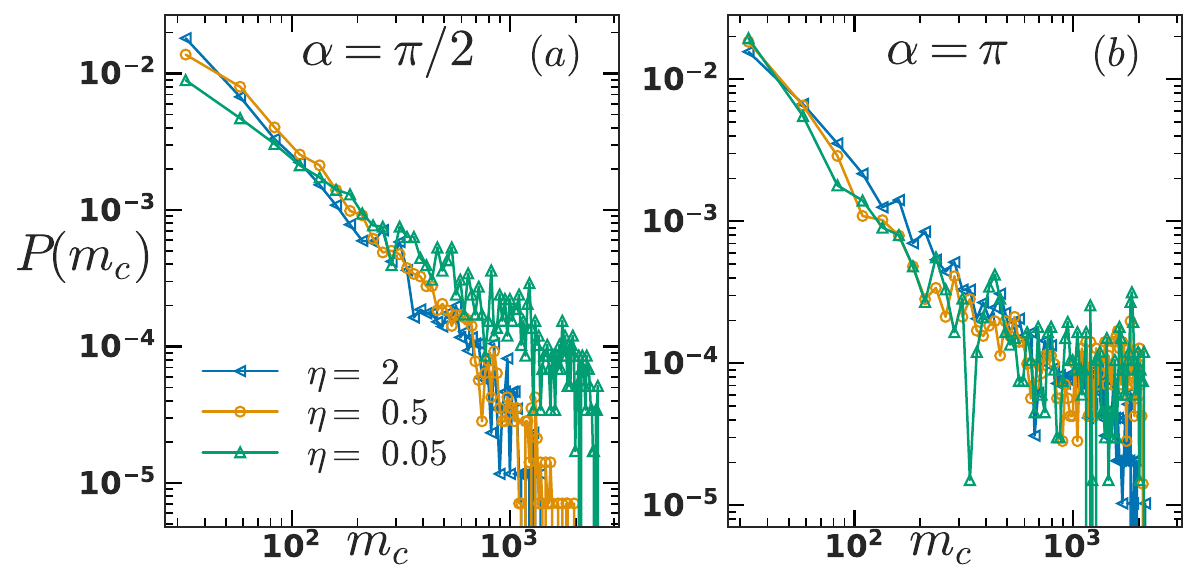}
            \caption{ Distribution of the cluster mass distribution $P(m_c)$ versus $m_c$ for (a) $\alpha=\pi/2$ and (b) $\pi$ for different values of noise $\eta$. All the data are calculated for $N=2560$ and $L=32$.  }
            \label{dist_clstmass_appendix}
        \end{figure}
        As seen before flocking becomes difficult with decreasing  value of $\alpha$. Fig.~\ref{dist_clstmass_appendix}(a)-(b) shows the normalized cluster mass distribution for   $\alpha=\pi/2$ and $\pi$. For $\alpha=\pi/2$ the behavior of $P(m_c)$ for different noises are similar to that for $\alpha=\pi/4$. The probability of larger clusters decreases with increasing $\eta$. Conversely, for $\alpha=\pi$ the distributions for all values of $\eta$ more or less fall on each other suggesting similar probability of formation of  larger clusters.

       \section{Time evolution at intermediate vision-angle $\alpha=\pi/2$ with $\eta=0.5$} \label{Appendix_snap_alpha_pi2}
        \begin{figure}[htb]
            \includegraphics[width=8.5cm, height=3.0cm]{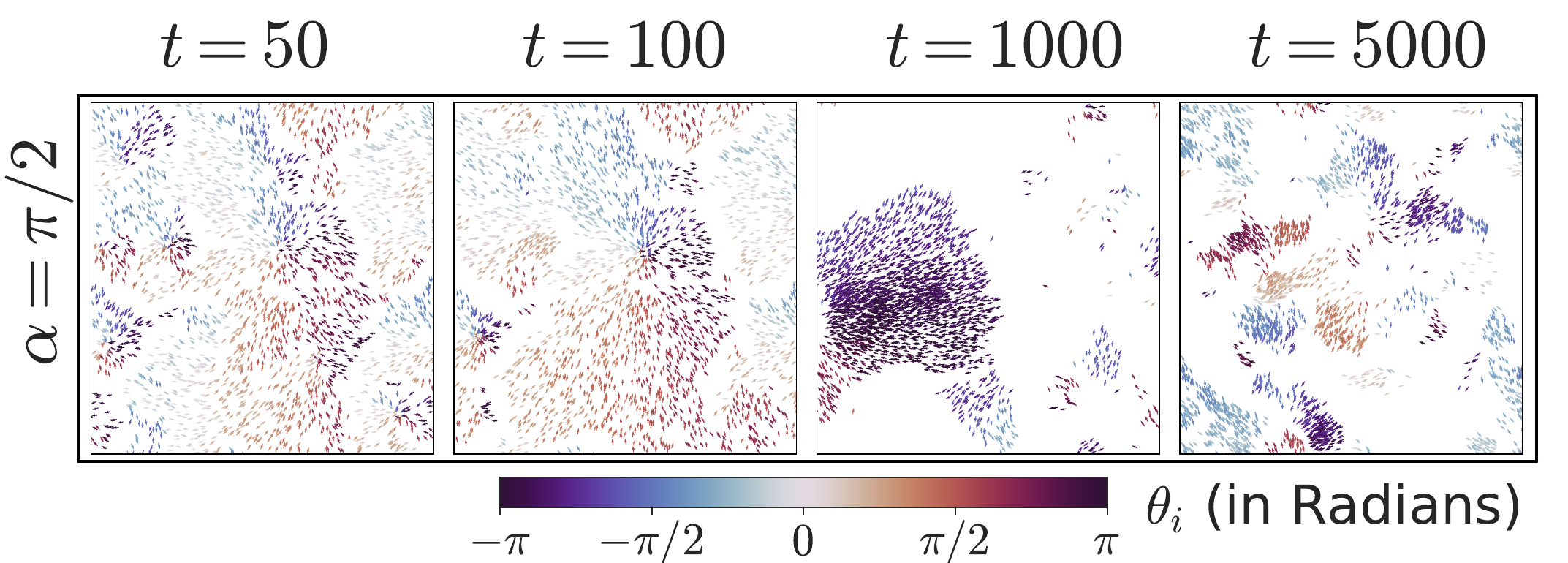}
            \caption{Time-evolution of the velocity field for vision-angle $\alpha=\pi/2$ for the medium noise $\eta=0.5$. Different colors identify particles with their respective  orientations. The snapshots are with  $N=2560$ for a system size $L=32$. }
            \label{time_evolution_medium_noise}
        \end{figure}
       
        Typical time evolution snapshots corresponding to $\alpha=\pi/2$ and $\eta=0.5$ shows the clustering and fragmentation events in a better manner. Starting from a uniform distribution of particles a larger cluster forms at $t=1000$. However, at much later time this breaks and forms a few smaller ones with local ordering of velocities within them. For these parameters $\langle v_a \rangle$ saturates at $\approx 0.5$.
        
        \section{Time evolution at high noise $\eta=2.0$} \label{Appendix_snap_time}
        
        \begin{figure}[htb]
            \includegraphics[width=8.5cm, height=4.6cm]{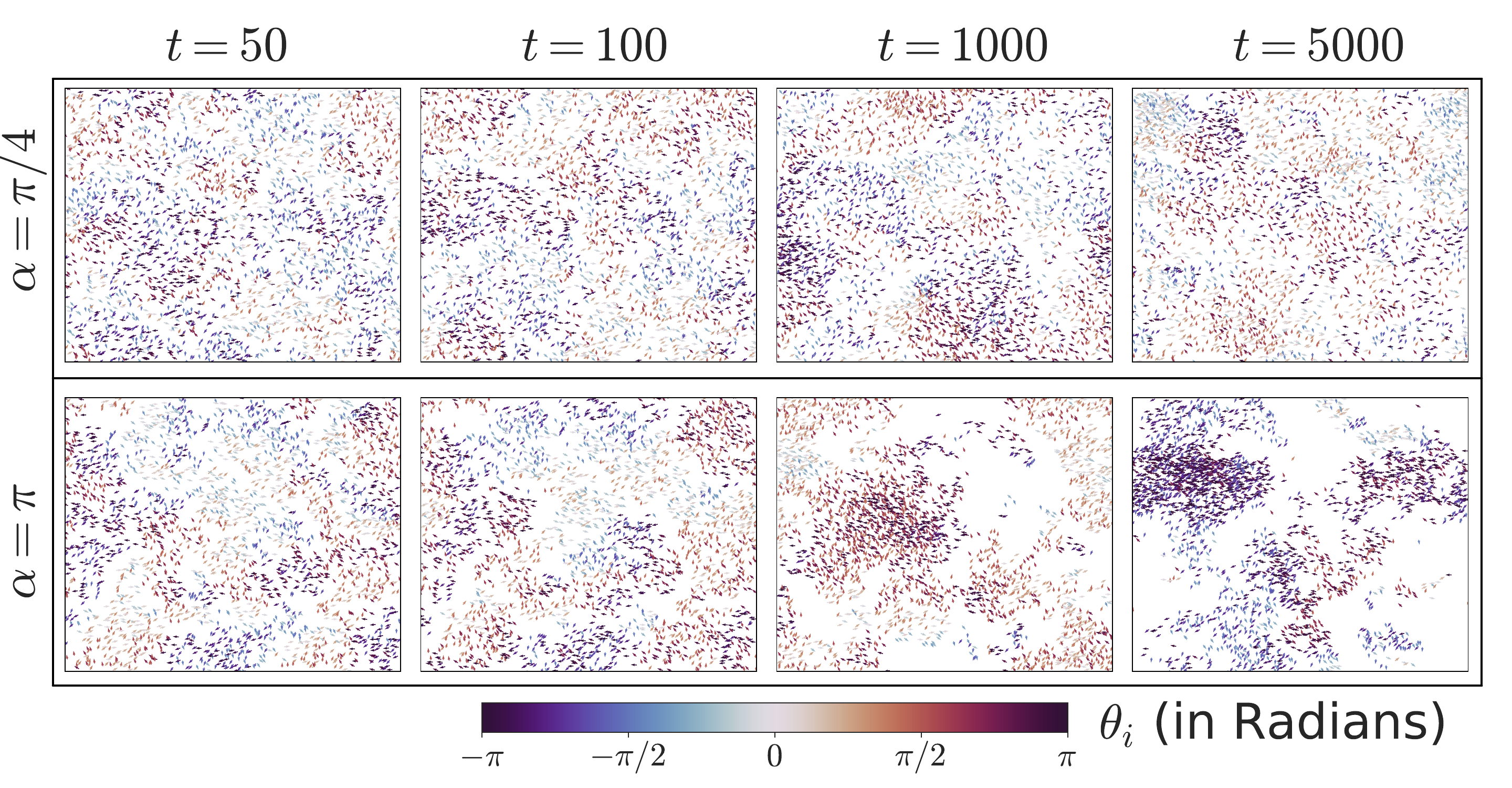}
            \caption{Time-evolution  snapshots of the velocity field for vision-angle $\alpha=\pi/4$ and $\pi$ for the higher noise $\eta=2.0$. Different colors identify particles with their respective  orientations. All the snapshots are with  $N=2560$ for a system size $L=32$. }
            \label{time_evolution_high_noise}
        \end{figure}       
        At high noise, i.e., $\eta=2.0$ time evolution snapshots of the velocities  are presented in Fig~\ref{time_evolution_high_noise} for $\alpha=\pi/4$ and $\pi$. Global flocking is difficult at high noise as also seen from the value of $
        \langle v_a \rangle$ in Fig~\ref{va_heatmap}. For $\alpha=\pi$  spatial clustering is somewhat visible with their local velocity ordering. However, for $\alpha=\pi/4$ the system remains homogeneous without any velocity ordering ($\langle v_a \rangle \approx 0$) or particle clustering.  
        \section{Time evolution of CCF}\label{Appendix_ccf}
        \begin{figure}[htb]
            \includegraphics[width=8.2cm, height=6.8cm]{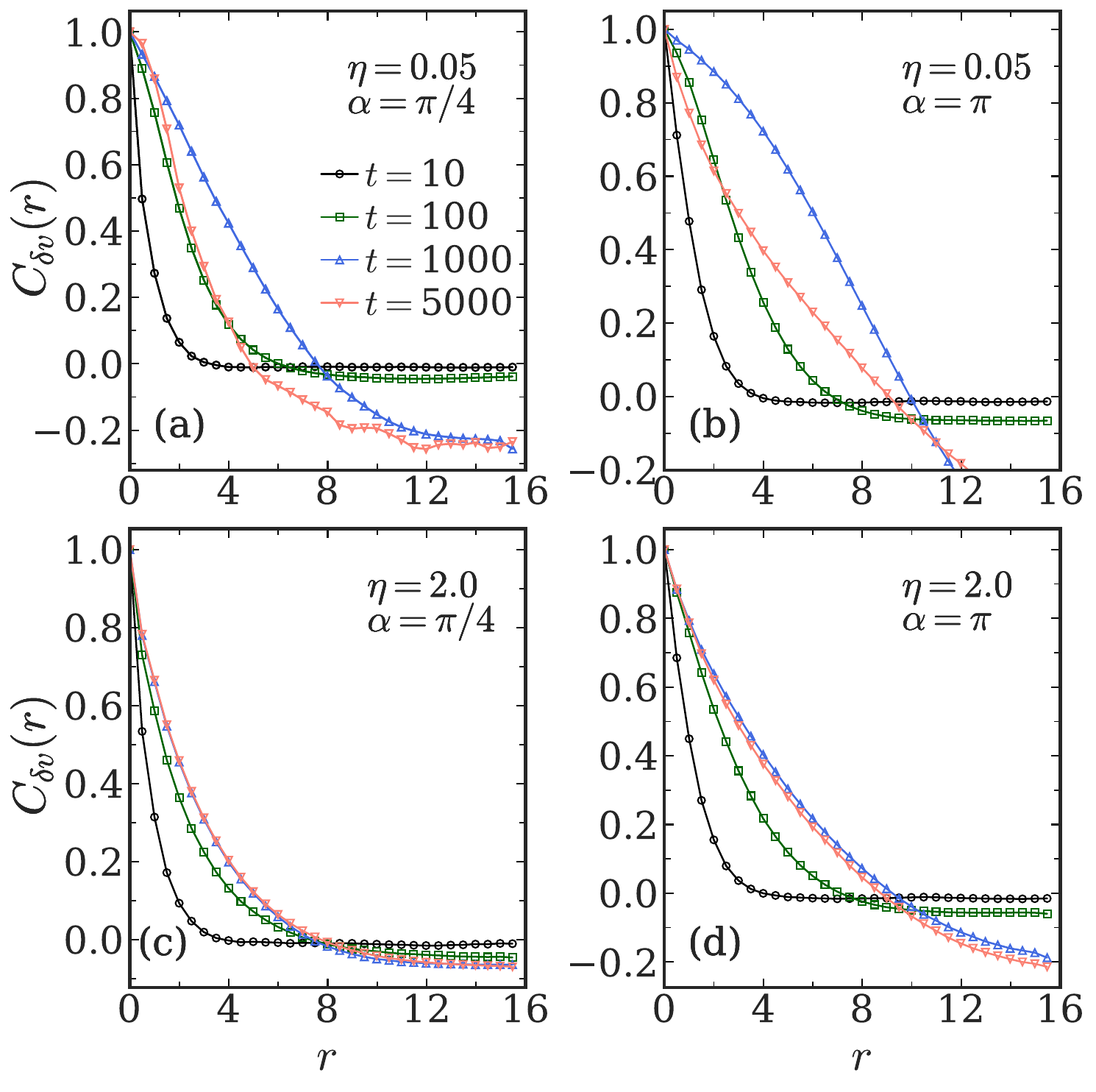}
            \caption{Time evolution of normalized connected correlation function $C_{\delta v}(r,t)$ as a function of $r$, for a few combinations of $\alpha$ and $\eta$, as mentioned in (a)-(d). The corresponding times are mentioned in frame (a). All data are for $L=32$ with $N=2560$.}
            \label{ccf-figure}
        \end{figure}
        In Fig.~\ref{ccf-figure} we show time evolution of the CCF for the cases ($\eta=0.05$ and $2.0$ with $\alpha=\pi/4$ and $\pi$), as considered for $C_v(r,t)$ in Fig.~\ref{velo_crl_time}. Like in $C_v(r,t)$, here also, decay of $C_{\delta v}(r,t)$ becomes slower with time, suggesting development of correlation in the fluctuations as well. However, for the lower noise case for both values of $\alpha$, decay of $C_{\delta v}(r,t)$ with time show non-monotonic behavior. A faster decay at very late time (data for $t=5000$) indicates the reduction of the length-scale associated with fluctuation correlations similar to what is observed for the VCF as well in Fig.~\ref{velo_crl_time}. As mentioned, this can be due to increase of local density due to lowering of interparticle separation, or due to fragmentation of clusters. However, for higher noises such non-monotonic behavior in time evolution of $C_{\delta v}(r,t)$ with time is absent.

        \section{Length scale calculated from CCF}\label{Appendix_correl_length} 
\begin{figure}[t!]
            \includegraphics[width=7.9cm, height=7.6cm]{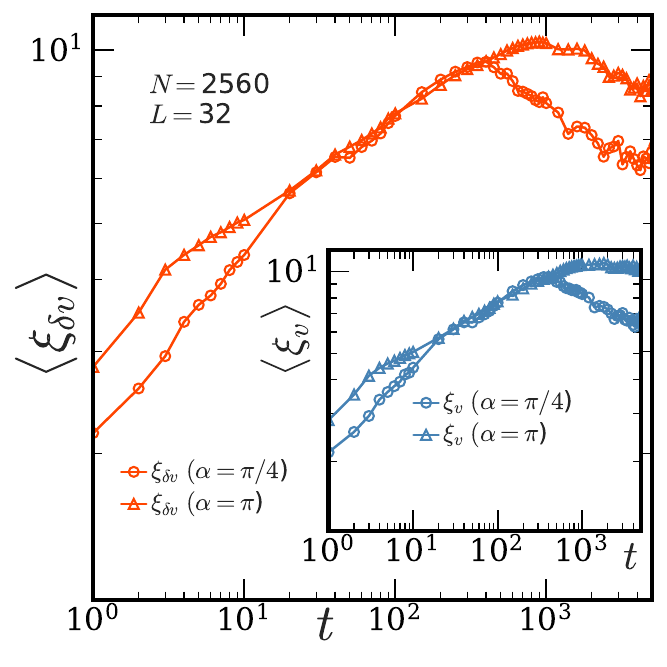}
            \caption{Length-scale $\langle \xi_{\delta v}(t)\rangle$ calculated from the decay of $C_{\delta v}(r,t)$ versus $t$, for $\alpha=\pi/4$ and $\pi$ with $\eta=0.5$. Comparison plots for the velocity correlation length $\xi_v(t)$ is shown in the inset. All the data are for $N=2560$ for $L=32$.}
            \label{connectedcorrlength_time}
        \end{figure}

        Since the deviation $\delta v_i$ provides idea of the behavior related to fluctuations with respect to the global average, the time evolution of the connected correlation length $\langle \xi_{\delta v}\rangle$ can provide a better insight regarding velocity coherence. In other words, $\xi_{\delta v}$ estimates the length-scale over which fluctuations are correlated, providing a more accurate picture distinguishing between global and local orders. Similar to $\xi_v$, here, $\xi_{\delta v}$ is calculated from the decay of $C_{\delta v}(r,t)$ shown in Fig.~\ref{ccf-figure}. We plot the time evolution of $\langle \xi_{\delta v}\rangle$ in Fig~\ref{connectedcorrlength_time}  for $\alpha=\pi/4$ and $\pi$. Inset shows the corresponding data for $\langle \xi_v\rangle$ for comparison. Starting from a random state, the development of correlation appears to be similar manner for the particle velocities as well as its fluctuation. However, at late times, in the steady-state, the behavior seems different for different vision angles. Decrease in $\langle \xi_v \rangle$ due to noise or fragmentation of clusters definitely  corresponds to a decrease in $\langle \xi_{\delta v} \rangle$. This is also confirmed from the presented data for $\alpha=\pi/4$. However,  interestingly, for $\alpha=\pi$, even though $\langle \xi_v \rangle$ saturates and remain constant with time, $\langle \xi_{\delta v} \rangle$ decreases. This suggests that even in the presence of an average global direction of flocking, the fluctuations for different clusters can be uncorrelated. This is also consistent with  the spread of $\delta \vec{v}_i$, as seen from Fig.~\ref{dist_velo_field}(f). This also indicates that in presence of noise the particle directions may be uncorrelated within the cluster as well despite of being part of the same cluster due to an effective cohesive nature among the particles.
        
        %\section{ Appendix F: Plots for $C_v(r)$ in log-log }\label{Appendix_vcf_log}
        
        %\begin{figure}
        %    \includegraphics[width=1.0\linewidth]{fig_vcf_app.pdf}
        %    \caption{Log-log plot of the velocity correlation function $C_v(r)$ vs $r$ for $\eta=0.05$ and various value of $\alpha$. All the data is with $N=2560$ for a system size $L=32$ and has been averaged over 100 different initial configurations.}
        %   

     %   \bibliography{ref.bib}
     
     %

        \end{document}